\documentclass[%
 preprint,prx,
amsmath,amssymb,
superscriptaddress,
aps, 11pt]{revtex4-1}
\usepackage[nottoc]{tocbibind}
\settocbibname{References}
\usepackage[version=4]{mhchem} 

\usepackage{amsmath}
\usepackage{amssymb}
\usepackage{dcolumn}
\usepackage{bm}
\usepackage{hyperref}
\usepackage{verbatim}
\usepackage{color}
\usepackage{xcolor}
\usepackage{mhchem}
\usepackage{url}

\usepackage{verbatim}
\usepackage{graphics}
\usepackage{lipsum}% http://ctan.org/pkg/lipsum
\usepackage{graphicx}% http://ctan.org/pkg/graphicx

\usepackage{subfigure}
\usepackage{subcaption}
\usepackage{wrapfig}
\usepackage{epsfig}
\usepackage{float}
\usepackage{array}
\usepackage{psfrag}
\usepackage{color}
\usepackage{bbold}

\newtheorem{theorem}{Theorem}[section]%added by Shuang 

\definecolor{azure}{rgb}{0.0, 0.5, 1.0}
\definecolor{asparagus}{rgb}{0.53, 0.66, 0.42}
\definecolor{ballblue}{rgb}{0.13, 0.67, 0.8}
\definecolor{sgreen}{rgb}{0.0, 0.8, 0.35}
\definecolor{darkgreen}{rgb}{0.0, 0.5, 0.0}
\definecolor{sred}{rgb}{0.9, 0.6, 0.4}
\usepackage{textcomp}
\usepackage{tikz}

\newcommand{\vect}[1]{\boldsymbol{#1}}

\usetikzlibrary{shapes.misc}
\tikzset{cross/.style={cross out, draw=black, minimum size=2*(#1-\pgflinewidth), inner sep=0pt, outer sep=0pt}, cross/.default={1pt}}
\usepackage[]{color}

\usepackage{caption}

\begin{document}

%%%
\title{
A ternary mixture model with dynamic boundary conditions
}
%%%
\begin{abstract}
%write the structure firstly and add the contexts.
\end{abstract}

%%%%%%
\author{{Shuang Liu}}
\affiliation{Department of Mathematics,  University of North Texas, {1155 Union Circle, Denton, Texas 76203-5017,} USA}
\author{{Yue Wu}}
\affiliation{{Department of Mathematical Sciences}, University of Nottingham Ningbo China, {Taikang East Road 199, 315100 Ningbo, China}}
\author{{Xueping Zhao}
\footnote{\label{CA}Corresponding author: {{Xueping.Zhao@nottingham.edu.cn}}}}
\affiliation{{Department of Mathematical Sciences}, University of Nottingham Ningbo China, {Taikang East Road 199, 315100 Ningbo, China}}
%%%%%%

% \date{\today}

\begin{abstract}
% The question of how the multi-phase multi-component mixture is influenced by the {short-range} interaction with the solid wall in a confined geometry plays an important role in life science and engineering. 
% To our knowledge, we firstly, in this work, extend the Cahn-Hilliard  model with dynamic boundary conditions for a binary mixture to a ternary mixture based on the Onsager principle, which considers cross-coupling relation between forces and fluxes in both bulk and surface. 
% Moreover, we develop a linear, second-order, unconditionally energy stable numerical scheme to solve the governing equation system based on the Invariant Energy Quadratization method. Utilizing the efficient numerical solver, we investigate the effects of wall-mixture interaction and dynamic boundary conditions on the multi-component multi-phase mixture, e.g., spontaneous phase separation and coarsening processes, wettability of droplet on the surface. 
% We find that the wall-mixture interaction not only controls the surface phenomena, such as the contact angle of the droplets wetting on the surface, but also the patterns inside the bulk far away from the surface. 
% Furthermore, cross-coupling relaxation parameter in the kinetic model make an influence on the coarsening patterns and the droplet spreading processes on the solid surface.  
% Our work provides a general framework to study the multi-component mixture in confined geometry.

{The influence of short-range interactions between a multi-phase, multi-component mixture and a solid wall in confined geometries is crucial in life sciences and engineering. In this work, we extend the Cahn-Hilliard model with dynamic boundary conditions from a binary to a ternary mixture, employing the Onsager principle that accounts for cross-coupling between forces and fluxes in both bulk and surface. Moreover, we have developed a linear, second-order, and unconditionally energy stable numerical scheme for solving the governing equations, utilizing the Invariant Energy Quadratization (IEQ) method. This efficient solver allows us to explore the impacts of wall-mixture interactions and dynamic boundary conditions on phenomena like spontaneous phase separation, coarsening processes, and the wettability of droplets on surfaces. We observe that wall-mixture interactions influence not only surface phenomena, such as droplet contact angles, but also patterns deep within the bulk. Additionally, the relaxation rates control the droplet spreading on surfaces. Furthermore, the cross-coupling relaxation rates in the bulk significantly affect coarsening patterns. Our work establishes a comprehensive framework for studying multi-component mixtures in confined geometries.}
% \SL{Two "Furthermore" in the abstract}
\end{abstract}

\maketitle
% \tableofcontents

\section{Introduction}
% to dos
% outlines for paragraphs
% first/last sentence
% add references.
% further literatue survey in google. add one after getting one. 

% Think and writting. finish the introduction today. 
% use grammar to double check the grammar. 

% 1. biological/chemical background/application of liquid condensate wetting. The question of how the process of spontaneous phase separation in binary mixtures (spinodal decomposition; see, e.g., [1,2]) is influenced by the presence of walls has gained much attention recently both in experimental [3– 8] and theoretical work [9 – 15]. Also, biomolecule condensates wetting on the membrane gain a lot of attentions. 
% last sentence: significance of modeling. 
% % (I want to add the significance of modeling, but I do not know exactly in my mind. I can add similar points in the draft initially. Later on, I can find my evidence for my points.  )
% phase separation in the presence of confining walls.
% phase separation in the presence of biological lipid membranes in living cells. 

% first paragraph outlines of sentences. 
% \com{Key words: three phases coexisting, compound droplet wetting behavior, weighted contact angel, extra phase field for the solid phase}

The physics of multi-component mixtures, including spontaneous phase separation, nucleation and growth, and coarsening have been well-studied both theoretically and experimentally. The question of how the multi-component multi-phase system {interacts} with sold walls in confined {geometries} has gained much attention due to its wide applications in life science and engineering{,} e.g. membrane-less organelles formation and wetting on the membrane-bound organelles in living cells\cite{Brangwynne2009, Gall1999, Brangwynne2009,
Knorr_wetting_nature_2021}; micro-fluid close to a solid interface in microfluidic devices \cite{Tabeling_2005}; thin films of polymer blends in a slab geometry \cite{PRL_Bruder_1992, PRL_Krausch_1993, PRL_Sung_1996}{, and so on.}

{Numerous methods have been developed to study interface problems and their interaction with solid walls. For instance, the lattice Boltzmann model has garnered considerable attention for simulating multi-phase and multi-component systems with complex boundary conditions, as demonstrated by several studies \cite{PRA_LBM_Gunstensen_1991, PRE_LBM_Shan_Chen_1993, PRE_LBM_Shan_Chen_1994, PRL_LBM_Swift_1995, PRE_LBM_Swift_1996}. Additionally, the exploration of wetting properties in ternary mixture models has been a significant focus \cite{schmieschek_harting_2011, Yan_energies_2021, Yan_JRMGE_2022}. For example, Liang et al. \cite{LIANG2019} formulated a wetting boundary condition for ternary fluids interacting with a solid substrate, applying this to the lattice Boltzmann model. 
The Volume Of Fluid (VOF) technique is particularly effective in scenarios involving solid obstacles, as it integrates the contact angle by applying appropriate boundary conditions to the fluid-function's gradient at solid interfaces. Therefore, VOF has been extensively used in fluid dynamics analyses within complex porous structures \cite{Ferrari_AWR_2013,Ferrari_AWR_2014,Ferrari_WRR_2015,Li_PRE_2018,Li_PRF_2021}. However, among all these methods, the phase field method stands out as the most effective tool for modeling and simulating multi-phase systems. Its inherent flexibility in handling complex topological changes in the phases, coupled with its ability to seamlessly integrate with various boundary conditions, makes it a superior choice for addressing the complexities of interface problems in multi-component systems. }

Cahn--Hilliard equation is a fundamental phase field model. Since it is firstly proposed in materials science to describe the pattern formation evolution of microstructures during the phase separation process in binary alloys \cite{Cahn_1958, CAHN1962}, the Cahn--Hilliard equation and its variants have been successfully applied in a wide variety of segregation-like phenomena in science, see for instance  \cite{FRIED1993, Bai_1995, GURTIN1996, PRL_1997_WDieterich, Dieterich_2001,  Goldstein2011, Liu-Wu_Model, KLLM_2021} and the references therein.
In this study, we denote the domain as $\Omega$, the boundary of the domain $\Gamma$. We introduce $\phi$, which represents the volume fraction of a specific component within the mixture. $\phi \in (0, 1)$ is dimensionless and quantifies the proportion of the total volume occupied by a given component, providing insight into the concentration and distribution of that component within the system. The total free energy of the system reads
\begin{align}
    F = \int_{\Omega} \Big[ f_b(\phi) + \frac{\kappa}{2} |\nabla \phi|^2\Big] d\vect{x},
\end{align}where $f_b(\phi)$ is the bulk free energy density function, the term $\frac{\kappa}{2} |\nabla \phi|^2$ represents the interface energy between phases.   
The Cahn--Hilliard equation has the following format:
{
\begin{align}
    \partial_t \phi = \nabla M \cdot \nabla  \mu  \, ,
\end{align}}where mobility $M$ can be either a constant or concentration-dependent function, and $\mu$ is the chemical potential of {the} component $\phi$. 
The boundary conditions include the following two categories:
\begin{enumerate}\setlength{\itemsep}{0pt}%
    \item periodic boundary conditions.
    \item physical boundary conditions. e.g. homogeneous Neumann boundary conditions.
\end{enumerate}
However, in some systems, e.g.
phase separation in confined geometries, the condensate will interplay with boundaries, and thus more generic boundary conditions, which describe the interaction between wall and condensates, are introduced.  
% [The significance/application of generic boundary conditions. ]
In this work, we will focus on the study of generic boundary conditions. 

% Third paragraph
The influence of boundaries (solid walls) on the phase separation process of binary mixtures has attracted {lots of} attention from scientists. For instance, W. Dieterich, etc. \cite{PRL_1997_WDieterich} {introduced} the dynamic boundary conditions in 1997:
\begin{subequations}\label{eq:DBC1_sum}
    \begin{align} \label{eq:DBC1}
    \partial_n \mu|_{\Gamma} &= 0,\\ \label{eq:DBC1-2}
    \partial_t \phi|_{\Gamma} & = - \Gamma_s \Big[ \kappa \, \partial_{n} \phi + {f_s}' - \kappa_{\Gamma} \Delta_{{\Gamma}} \phi_{\Gamma}   \Big], %\nonumber 
\end{align}
\end{subequations} {where} $f_s$ is the surface free energy density which represents the short-range interaction between mixture components and the solid wall{, $\Delta_{\Gamma}$ stands for the Laplace–Beltrami operator on the boundary surface $\Gamma$, and} $\Gamma_s$ defines a surface kinetic coefficient and the term $\kappa \, \partial_n \phi$ is due to the surface contribution that comes from the variation of the bulk free energy $f_b$. The boundary \eqref{eq:DBC1} is simply the condition that no current can flow through the surface, while the boundary \eqref{eq:DBC1-2} can be derived by requiring the system to tend  to minimize its surface free energy{,}
\begin{align}
    F_s = \int_{\Omega} \Big[ \frac{\kappa_{\Gamma}}{2} |\nabla_{\Gamma} \phi_{\Gamma}|^2 + f_s\Big].
\end{align}
Boundary conditions of similar form as Eqs.~\eqref{eq:DBC1_sum} have also been derived from a semi-infinite Ising model with Kawasaki spin exchange dynamics \cite{1991_Binder, PRE_1994_Puri}.
In 2011, Goldstein, etc. {derived} a Cahn--Hilliard model in a domain with non-permeable walls\cite{Goldstein2011}. In this model, they {assumed} that the total mass in the bulk and on the boundary, {i.e. 
\begin{align}
\int_\Omega \phi(\vect{x}) d\vect{x} + \int_{\Gamma} \phi(\vect{x}) dS,
\end{align}}is conserved. Its boundary conditions read 
\begin{subequations}\label{eq:DBC2}
\begin{align} 
   \partial_t \phi|_{\Gamma}  &=  \nabla_{\Gamma} M \cdot \nabla_{\Gamma} \mu_{\Gamma} - \beta  M  \partial_n \mu,\\
    \mu|_{\Gamma} &=  \beta  \Big[ \kappa \, \partial_{n} \phi + {f_s}' - \kappa_{\Gamma} \Delta_{{\Gamma}} \phi_{\Gamma} \Big] .
\end{align}
\end{subequations}
In addition, they {proved} the existence and uniqueness of weak solutions and {studied} their asymptotic behavior as time goes to infinity. 

% {double check the formula here.}
Liu Chun and Wu Hao \cite{Liu-Wu_Model} introduced a Liu-Wu model in 2019 with non-flux boundary conditions for chemical potential and Cahn--Hilliard type gradient flow on the boundary{, i.e.} 
\begin{subequations}\label{eq:DBC3}
    \begin{align} 
   \partial_n \mu & = 0,\\
    \partial_t \phi|_{\Gamma} &=  \nabla_{{\Gamma}} M \cdot \nabla_{{\Gamma}}   \Big[ \kappa \, \partial_{n} \phi + {f_s}' - \kappa_{\Gamma} \Delta_{{\Gamma}} \phi_{\Gamma} \Big] .
\end{align}
\end{subequations}
In 2021, Patrik Knopf and collaborators proposed a general model\cite{KLLM_2021}: KLLM model.
\begin{subequations}\label{eq:DBC4}
    \begin{align} 
    L\partial_{n} \mu & = -\mu|_{\Gamma}  + \beta  \Big[ \kappa \, \partial_{n} \phi + {f_s}' - \kappa_{\Gamma} \Delta_{{\Gamma}} \phi_{\Gamma} \Big] ,\\
    \partial_t \phi|_{\Gamma} &=  \nabla_{{\Gamma}}  M \cdot \nabla_{{\Gamma}}   \Big[ \kappa \, \partial_{n} \phi + {f_s}' - \kappa_{\Gamma} \Delta_{{\Gamma}} \phi_{\Gamma}   \Big] {-\beta M \partial_{n} \mu}.
\end{align}
\end{subequations}
Harald Garcke etc \cite{GARCKE2022} did the nonlinear {analysis, long-time dynamics} of the Cahn–Hilliard equation with kinetic rate dependent dynamic boundary conditions for the KLLM model. We refer to \cite{Wu_Hao_review_2022} for reviews of the model and analysis.

Correspondingly, there are numerical studies for binary mixture models with dynamic boundary conditions{. For example,} W. Dieterich etc. proposed an implicit numerical scheme \cite{Dieterich_2001} to solve the Cahn--Hilliard equation with \eqref{eq:DBC1_sum}. The authors discretized the solution in space using the finite difference method on edge points. {Stability} analysis {was} derived and a variable {time-stepping} strategy {was} employed to simulate the system for a long time. Furthermore, the numerical scheme is conditionally gradient stable, which means that the free energy decreases monotonously in time. 
In 2017, Shuji Yoshikawa and collaborators \cite{Yoshikawa_2017} constructed a structure-preserving finite difference scheme for the Cahn--Hilliard equation with dynamic boundary conditions in the one-dimensional case  for Goldstein model \eqref{eq:DBC2}. Unlike {the} space discretization in \cite{Dieterich_2001}, the author used the finite difference method on center points. {The existence of the solution and the error estimate were also obtained.
% We note that {the} discretization on the boundary is not exactly on the boundary line in this work. 
In addition, the laws of mass conservation and energy dissipation were satisfied in the discrete level.} Zhang Zhengru and collaborators {developed} a {second} order stabilized semi-implicit scheme \cite{Zhang_Zhengru_2022} for Liu-Wu model \eqref{eq:DBC3}. The corresponding energy stability and convergence analysis of the scheme {were} derived theoretically.

{Besides the binary mixture models, phase field models for ternary mixture systems {have} been developed due to their frequent real-world applications, as indicated in studies \cite{re1_1,re1_2,re1_3}. Wetting behavior\cite{3_2016,4_2023} {has} also been proposed based on the phase field models. For example, a phase field model for multi-component Cahn-Hilliard systems in complex domains was developed in \cite{3_2016}, considering contact angle or no mass flow boundary conditions. Building on this, \cite{4_2023} extended the geometrical formulation of wetting conditions using weighted contact angles defined within the implementation of wetting conditions, following the concept of the {diffusive} interface  \cite{WCA}. This innovative phase field model efficiently describes the dynamic behavior of compound droplets in contact with solid objects.}

{However, we note that dynamic boundary conditions are missing in the aforementioned multi-component phase field models. Moreover, to the best of our knowledge, a ternary mixture model with the force-flux cross-coupling relations has not been explored yet.} In summary, based on the above discussion, we focus on the following aims in this work:
\begin{enumerate}\setlength{\itemsep}{0pt}%
    \item Derive a model for a ternary mixture {with the force-flux cross-coupling relations} in a confined geometry with dynamic boundary conditions based on the Onsager principle and irrversible thermodynamics \cite{OnsagerL1, OnsagerL2,  Yang_Li_Forest_Wang2016, Liu_2008_Onsager, Doi_2021, Frank_JProst2008}.
    \item Develop a linear second-order unconditionally energy stable numerical scheme for the derived model based on the Invariant Energy Quadratization(IEQ) method \cite{Yang_Zhao_Wang_Shen2017, Yang-EQ-PFC, ZYLW_SISC_2016, Jia_XF_YZ_XP_XG_Jun_Q2018}.
    \item Investigate equilibrium and out of equilibrium properties of the system based on the numerical simulations. 
\end{enumerate}

% In section \ref{sec:modelderive}, a kinetic model for a ternary mixture in the confined geometry and interacting with the surrounding solid walls, is derived {remove[}based on the Onsager principle and{]} irreversible thermodynamics. The total mass is conserved. The total energy decreases with respect to time.   Then, we equivalently reformulate the mathematical model and develop the numerical scheme for the reformulated model based on the IEQ method in section \ref{sec:numericalscheme}.In section \ref{sec:parameterstudy},  the numerical studies for the dependence of spontaneous phase separation and wettability of the surface on the model parameters are presented. Finally, in section \ref{sec:conclusions}, we draw conclusions.

{Section \ref{sec:modelderive} details the derivation of a kinetic model. This model, which addresses a ternary mixture within confined geometries and its interaction with adjacent solid walls, is formulated using principles of irreversible thermodynamics. It features conservation of total mass and a decrease in total energy over time. The mathematical model undergoes an equivalent reformulation based on the IEQ method in Section \ref{sec:numericalscheme}, where we also introduce a second-order, linear, unconditional energy stable numerical scheme for the revised model. Section \ref{sec:parameterstudy} examines how spontaneous phase separation and surface wettability are influenced by various model parameters using numerical studies. The paper concludes in Section \ref{sec:conclusions}, where we summarize our findings.}

\section{Mathematical model derivation}\label{sec:modelderive}
In this section, we will extend the binary mixture model with dynamic boundary conditions to a ternary mixture model, based on the Onsager principles\cite{OnsagerL1, OnsagerL2, Yang_Li_Forest_Wang2016, Liu_2008_Onsager, Doi_2021}. 
We denote the volume fractions of three components as $\phi_1$, $\phi_2$, and $\phi_3$, where $\phi_3 = 1-\phi_1 - \phi_2$ based on the incompressibility assumption. The total free energy of the system includes two contributions: one is the free energy in the bulk $\Omega$, and another on the surface $\Gamma$, i.e.
\begin{align}
E = E_{\text{bulk}} + E_{\text{surf}},
\end{align}
where the bulk free energy $E_{\text{bulk}}$ and surface free energy $E_{\text{surface}}$ read
\begin{align}
E_{\text{bulk}} = \int_{\Omega} \Big[ f_b(\phi_1, \phi_2) + \frac{\kappa_1}{2}|\nabla \phi_1|^2 + \frac{\kappa_2}{2}|\nabla \phi_2|^2 + \kappa_{12} \nabla \phi_1 \cdot \nabla \phi_2 \Big],\\
E_{\text{surf}} = \int_{\Gamma} \Big[ f_s(\phi_{1 \Gamma}, \phi_{2 \Gamma}) + \frac{\kappa_{1\Gamma}}{2}|\nabla_{\Gamma} \phi_{1 \Gamma}|^2 + \frac{\kappa_{2\Gamma}}{2}|\nabla_{\Gamma} \phi_{2 \Gamma}|^2 + \kappa_{12\Gamma}\nabla_{\Gamma}\phi_{1\Gamma} \cdot \nabla_{\Gamma} \phi_{2\Gamma} \Big].
\end{align}
$\kappa_{i}$, $\kappa_{i\Gamma}$, $i = 1,2, 12$ are gradient coefficients associated with interface free energies. For simplicity, we set $\kappa_{12} = 0$, and $\kappa_{12\Gamma} = 0$ in the following study. $\nabla_{\Gamma}$ denotes the gradient operator on the surface.
The bulk free energy density function $f_b(\phi_1, \phi_2)$ and surface free energy density function $f_s(\phi_{1 \Gamma}, \phi_{2 \Gamma})$ are defined as Flory-Huggins free energy and a second order polynomial based on application in polymer field \cite{Cahn_wetting_1977},
\begin{align}
    f_b(\phi_1, \phi_2) & = \frac{k_B T}{\nu} \Big[ \frac{\phi_1}{n_1} \ln \phi_1 + \frac{\phi_2}{n_2} \ln \phi_2 + (1-\phi_1 - \phi_2) \ln(1-\phi_1 - \phi_2) + \chi_{12} \phi_1 \phi_2  \nonumber\\ & \quad + \chi_{13} \phi_1 (1-\phi_1 -\phi_2) + \chi_{23} \phi_2 (1-\phi_1 - \phi_2) + \omega_1 \phi_1 + \omega_2 \phi_2 \Big], \\
    {f_s(\phi_{1\Gamma}, \phi_{2\Gamma)}} & = \frac{k_B T}{\nu_s} { \Big[ h_1 \phi_{1\Gamma} + g_1 \phi_{1\Gamma}^2 + h_2 \phi_{2\Gamma} + g_2 \phi_{2\Gamma}^2 + \gamma \phi_{1\Gamma} \phi_{2\Gamma} \Big]},
    % f_s(\phi_1, \phi_2) & = \frac{k_B T}{\nu_s} \Big[ h_1 \phi_1 + g_1 \phi_1^2 + h_2 \phi_2 + g_2 \phi_2^2 + \gamma \phi_1 \phi_2\Big],
\end{align}
where $\nu_i = n_i \nu$ represents the molecule volume of component $i$, $i=1,2$, and $\nu$ the molecule volume of component 3. For simplicity, we assume the molecule volumes of each component are equivalent, i.e. $\nu_1 = \nu_2 = \nu$ and $n_1 = n_2 = 1$. $k_B$ is the boltzmann constant and $T$ is the temperature of the isotropic system. The term $\chi_{ij} \phi_i \phi_j$ refers to the interaction strength between the component $i$ and the component $j$. $\omega_i \phi_i$ denotes the internal free energy for component $i$ in the bulk $\Omega$. On the surface, $h_i\phi_i$, $i=1,2$ is the linear order term, which denotes the interaction proportional to the component volume fractions. The quadratic $g_i \phi_i^2$ depicts the molecule-molecule interactions of each components near by the wall, while $\gamma \phi_1 \phi_2$ describes the coupling interaction between components on the surface.  

We assume that there is no chemical reaction in the system, that is, each component of the mixture conserves the total mass. We write the conservation laws as follows:
\begin{align}\label{eq:conservation_laws}
    \partial_t \phi_1 +\nabla \cdot \vect{J_1} &= 0,\\
    \partial_t \phi_2 + \nabla \cdot \vect{J_2} &= 0.
\end{align}
\subsection{Irreversible thermodynamics}
We consider a closed system in this work. According to irreversible thermodynamics, the entropy product rate is inversely proportional to the energy change rate. 
\begin{align}
- T \partial_t S & = \frac{dE}{dt}  = \frac{\partial E}{\partial \phi_1} \frac{\partial \phi_1}{\partial t}+ \frac{\partial E}{\partial \phi_2} \frac{\partial \phi_2}{\partial t} 
+ \frac{\partial E}{\partial \phi_{1 \Gamma} }\frac{\partial \phi_{1 \Gamma}}{\partial t} 
+ \frac{\partial E}{\partial \phi_{2 \Gamma}} \frac{\partial \phi_{2 \Gamma}}{\partial t} \nonumber \\
& \quad  + \frac{\partial E}{\partial \nabla \phi_1} \frac{\partial \nabla \phi_1}{\partial t}+ \frac{\partial E}{\partial \nabla \phi_2} \frac{\partial \nabla \phi_2}{\partial t} 
+ \frac{\partial E}{\partial \nabla_{\Gamma} \phi_{1 \Gamma} }\frac{\partial \nabla_{\Gamma} \phi_{1 \Gamma}}{\partial t} 
+ \frac{\partial E}{\partial \nabla_{\Gamma} \phi_{2 \Gamma}} \frac{\partial \nabla_{\Gamma} \phi_{2 \Gamma}}{\partial t} \\
% & = \int_{\Omega} dx \Big[\frac{\partial f_b}{\partial \phi_1}\partial_t \phi_1 + \kappa_1 \nabla \phi_1 \cdot \nabla \partial_t \phi_1 + \frac{\partial f_b}{\partial \phi_2}\partial_t \phi_2 + \kappa_2 \nabla \phi_2 \cdot \nabla \partial_t \phi_2 \Big]\nonumber \\
% & \quad + \int_{\Gamma} dS \Big[ \frac{\partial f_s}{\partial \phi_{1 \Gamma}} \partial_t \phi_{1\Gamma} + \kappa_{1\Gamma} \nabla \phi_{1\Gamma} \cdot \nabla \partial_t \phi_{1\Gamma} + \frac{\partial f_s}{\partial \phi_{2 \Gamma}} \partial_t \phi_{2\Gamma} + \kappa_{2\Gamma} \nabla \phi_{2\Gamma} \cdot \nabla \partial_t \phi_{2\Gamma} \Big]\nonumber\\
% & = \int_{\Omega} dx \Big[ (\frac{\partial f_b}{\partial \phi_1} - \kappa_1 \Delta \phi_1)\partial_t \phi_1 + (\frac{\partial f_b}{\partial \phi_2} - \kappa_2 \Delta \phi_2 )\partial_t \phi_2 \Big] + \int_{\Gamma} dS \Big[ (\vect{n}\cdot\kappa_1\nabla\phi_1 )\partial_t\phi_1+ (\vect{n}\cdot\kappa_2\nabla\phi_2)\partial_t\phi_2 \Big] \nonumber \\
% & \quad + \int_{\Gamma} dS \Big[(\frac{\partial f_s}{\partial \phi_{1 \Gamma}} -\kappa_{1 \Gamma} \Delta \phi_{1\Gamma}) \Big] \frac{\partial \phi_{1\Gamma}}{\partial t} + \int_{\Gamma} dS \Big[(\frac{\partial f_s}{\partial \phi_{2 \Gamma}} -\kappa_{2 \Gamma} \Delta \phi_{2\Gamma}) \Big] \frac{\partial \phi_{2\Gamma} }{\partial t} \nonumber\\
& = \int_{\Gamma} dS \Big[  \vect{n}\cdot \kappa _1 \nabla \phi_1  +  (\frac{\partial f_s}{\partial \phi_{1 \Gamma}} -\kappa_{1 \Gamma} \Delta_{\Gamma} \phi_1) \Big] \frac{\partial \phi_{1\Gamma}}{\partial t}\nonumber \\
& \quad + \int_{\Gamma} dS \Big[  \vect{n}\cdot \kappa_2 \nabla \phi_2  +  (\frac{\partial f_s}{\partial \phi_{2 \Gamma}} -\kappa_{2 \Gamma} \Delta_{\Gamma} \phi_2) \Big] \frac{\partial \phi_{2\Gamma} }{\partial t} \nonumber\\
& \quad - \int_{\Gamma} dS \Big[\frac{\mu_1}{\nu_1} \vect{n} \cdot \vect{J_1} + \frac{\mu_2}{\nu_2} \vect{n} \cdot \vect{J_2}\Big]  + \int_{\Omega} d\vect{x} \Big[\nabla \frac{\mu_1}{\nu_1}  \cdot \vect{J_1} + \nabla \frac{\mu_2}{\nu_2} \cdot \vect{J_2}\Big],
\end{align}where the chemical potentials in the bulk for components 1 and 2 are defined as
\begin{align}
    \mu_1 &= \nu_1 \frac{\delta E}{\delta \phi_1} = \nu_1 \Big[\frac{\partial f_b}{\partial \phi_1} - \kappa_1 \Delta \phi_1 \Big],\\
    \mu_2 &=  \nu_2 \frac{\delta E}{\delta \phi_2} = \nu_2 \Big[\frac{\partial f_b}{\partial \phi_2} - \kappa_2 \Delta \phi_2 \Big].
\end{align}
We achieve the conjugate fluxes and forces as follows:
\begin{subequations}
\begin{align}
\vect{J_1} & \longleftrightarrow  - \nabla\mu_1\, , & x \in \Omega\, , 
\\ 
\vect{J_2} & \longleftrightarrow  - \nabla\mu_2\, , & x \in \Omega\, , 
\\
\partial_t \phi_{1 \Gamma} & \longleftrightarrow -  \left(  \vect{n}\cdot \kappa _1 \nabla \phi_1  +  \frac{\partial f_s}{\partial \phi_{1 \Gamma}} -\kappa_{1 \Gamma} \Delta_{\Gamma} \phi_1 \right) \,,  & x \in \Gamma \,, 
\\ 
\partial_t \phi_{2 \Gamma} & \longleftrightarrow -  \left(  \vect{n}\cdot \kappa _2 \nabla \phi_2  +  \frac{\partial f_s}{\partial \phi_{2 \Gamma}} -\kappa_{2 \Gamma} \Delta_{\Gamma} \phi_2 \right) \,,  & x \in \Gamma \,, 
\end{align}
\end{subequations}

{Based on the Onsager principle, we define a symmetric positive definite mobility function between conjugate fluxes and forces in both the bulk $\Omega$ and the surface $\Gamma$, i.e.}
\begin{subequations}
\begin{align}
    \vect{J_1} & = - M_1(\phi_1, \phi_2)  \nabla \mu_1 - M_{12}(\phi_1, \phi_2)  \nabla \mu_2 , \\
     \vect{J_2} &=  - M_{12}(\phi_1, \phi_2)  \nabla \mu_1  - M_2(\phi_1, \phi_2)  \nabla \mu_2 ,\\
   \partial_t \phi_{1 \Gamma} &= -\Gamma_1 \Big[\vect{n} \cdot \kappa_1 \nabla \phi_1 - \kappa_{1 \Gamma} \Delta_{\Gamma} \phi_{1 \Gamma} + \frac{\partial f_s}{\partial \phi_{1 \Gamma}} \Big] - \Gamma_{12} \Big[\vect{n} \cdot \kappa_2 \nabla \phi_2 - \kappa_{2 \Gamma} \Delta_{\Gamma} \phi_{2 \Gamma} + \frac{\partial f_s}{\partial \phi_{2 \Gamma}} \Big]  , \\
   \partial_t \phi_{2 \Gamma} &=  - \Gamma_{12} \Big[\vect{n} \cdot \kappa_1 \nabla \phi_1 - \kappa_{1 \Gamma} \Delta_{\Gamma} \phi_{1 \Gamma} + \frac{\partial f_s}{\partial \phi_{1 \Gamma}} \Big] -\Gamma_2 \Big[\vect{n} \cdot \kappa_2 \nabla \phi_2 - \kappa_{2 \Gamma} \Delta_{\Gamma} \phi_{2 \Gamma} + \frac{\partial f_s}{\partial \phi_{2 \Gamma}}\Big] ,
\end{align}
\end{subequations}
where $\Gamma_1$, $\Gamma_2$ and $\Gamma_{12}$ are the relaxation parameters of dynamic boundary conditions, $M_{i}(\phi_1,\phi_2)$ and $M_{12}(\phi_1,\phi_2)$ are defined as 
\begin{align}
    M_1(\phi_1, \phi_2) &= m_1 \phi_1 (1 - \phi_1 - \phi_2), \\
    M_2(\phi_1, \phi_2) &= m_2 \phi_2 (1 - \phi_1 - \phi_2), \\
    M_{12}(\phi_1, \phi_2) &= m_{12} \phi_1 \phi_2 (1 - \phi_1 - \phi_2).
\end{align}
to remove the singularity from Flory-Huggins free energy density functions. 
$M_{i}(\phi_1,\phi_2)$ and $\Gamma_{i}$, $i = 1,2$ are the diagonal elements in the mobility matrix, depicting the direct effects of forces on the corresponding fluxes, while $M_{12}(\phi_1,\phi_2)$ and $\Gamma_{12}$ are the cross coupling terms denoting the cross interplay of forces and fluxes. The cross coupling relaxation rate refers to the rate at which perturbations in one
variable are coupled to changes in another variable in the system. We note that our model is an extension of the binary model developed in \cite{PRL_1997_WDieterich}. Our work provides a general framework that can extend all existing binary mixture models \cite{Wu_Hao_review_2022, Goldstein2011, Liu-Wu_Model, KLLM_2021} with dynamic boundary conditions to a ternary model. 

Combining the definition of fluxes achieved from the Onsager principle with the conservation laws \eqref{eq:conservation_laws}, we obtain the governing equations of the system as follows.
% \SL{several places I felt strange,  In 39, do we need $\kappa_{2\Gamma}$ instead of $\kappa_{1\Gamma}$? }:
\begin{subequations}
\begin{align}\
    \partial_t \phi_1 &= \nabla  M_1(\phi_1, \phi_2) {\cdot} \nabla \mu_1 + \nabla  M_{12}(\phi_1, \phi_2) {\cdot} \nabla \mu_2 , \\
    \partial_t \phi_2 &= \nabla  M_2(\phi_1, \phi_2) {\cdot} \nabla \mu_2 + \nabla  M_{12}(\phi_1, \phi_2) {\cdot} \nabla \mu_1,
    \end{align}
\end{subequations}
with boundary conditions:
\begin{subequations}\label{eq:dbc_derived1}
    \begin{align}
\partial_t \phi_{1 \Gamma} &= -\Gamma_1 \Big[\vect{n} \cdot \kappa_1 \nabla \phi_1 - \kappa_{1 \Gamma} \Delta_{\Gamma} \phi_{1 \Gamma} + \frac{\partial f_s}{\partial \phi_{1 \Gamma}} \Big] - \Gamma_{12} \Big[\vect{n} \cdot \kappa_2 \nabla \phi_2 - \kappa_{2 \Gamma} \Delta_{\Gamma} \phi_{2 \Gamma} + \frac{\partial f_s}{\partial \phi_{2 \Gamma}} \Big]  , \\
   \partial_t \phi_{2 \Gamma} &= -\Gamma_2 \Big[\vect{n} \cdot \kappa_2 \nabla \phi_2 - \kappa_{2 \Gamma} \Delta_{\Gamma} \phi_{2 \Gamma} + \frac{\partial f_s}{\partial \phi_{2 \Gamma}}\Big] - \Gamma_{12} \Big[\vect{n} \cdot \kappa_1 \nabla \phi_1 - \kappa_{1 \Gamma} \Delta_{\Gamma} \phi_{1 \Gamma} + \frac{\partial f_s}{\partial \phi_{1 \Gamma}} \Big],\\
\partial_{\vect{n}} \mu_1|_{\Gamma} & = 0, \\
\partial_{\vect{n}} \mu_2|_{\Gamma} & = 0.
\end{align}
\end{subequations}
We denote the diffusion coefficients as $D_i = m_i k_B T$, $i=1,2$, and $D_{12} = m_{12} k_B T$. In this work, we use the dynamic boundary condition \eqref{eq:dbc_derived1} in $x$-direction and the periodic boundary condition for all variables in $y$-direction. 

Our model has the following properties:
% \com{Shuang: update the mass conservation laws below to the format of theorem in latex.}\SL{like following?}
\begin{theorem}
The total mass of each component in the domain $\Omega$ is conserved, i.e. 
{
\begin{align}\label{eq:tmass2.1}
    \int_{\Omega} \phi_i(x,y,t) d\vect{x} =  \int_{\Omega} \phi_i(x,y,0) d\vect{x}, \quad i=1,2.
\end{align}
}
\end{theorem}
\textbf{proof:}
For $i=1$,
{
\begin{align}
    \frac{d}{dt}\int_{\Omega} \phi_1 dx &= \int_{\Omega} \partial_{t} \phi_1 dx= -\int_{\Omega} \nabla \cdot \vect{J_1} dx = \int_{\Gamma} \vect{J_1}\cdot\vect{n} dS\nonumber \\
    &= \int_{\Gamma} \Big[- M_1(\phi_1, \phi_2)  \nabla \mu_1 - M_{12}(\phi_1, \phi_2)  \nabla \mu_2\Big]\cdot\vect{n} dS = 0,
\end{align}}
as boundary condition $\partial_{\vect{n}} \mu_1|_{\Gamma}  = 0$ and $\partial_{\vect{n}} \mu_2|_{\Gamma} = 0$. Similar for $i=2$. 
Thus Eq.~\eqref{eq:tmass2.1} is obtained. 

\begin{theorem}
The total free energy is dissipative, i.e.
\begin{align}
    \partial_t E = \partial_t (E_{\text{\rm surf}} + E_{\text{\rm bulk}}) \leq  0.
\end{align}
\end{theorem}
\textbf{proof:}
For simplicity, we assume $\nu_1 = \nu_2 = \nu$, and we define
\begin{align}
    \mu_{1\Gamma} = \vect{n} \cdot \kappa_1 \nabla \phi_1 - \kappa_{1 \Gamma} \Delta_{\Gamma} \phi_{1 \Gamma} + \frac{\partial f_s}{\partial \phi_{1 \Gamma}},\\
    \mu_{2\Gamma} = \vect{n} \cdot \kappa_2 \nabla \phi_2 - \kappa_{2 \Gamma} \Delta_{\Gamma} \phi_{2 \Gamma} + \frac{\partial f_s}{\partial \phi_{2 \Gamma}}.
\end{align}
The total energy dissipation rate reads
\begin{align}
\partial_t E &= \int_{\Omega} dx \Big[ (\frac{\partial f_b}{\partial \phi_1} - \kappa_1 \Delta \phi_1)\partial_t \phi_1 + (\frac{\partial f_b}{\partial \phi_2} - \kappa_2 \Delta \phi_2 )\partial_t \phi_2 \Big]\nonumber \\
& \quad + \int_{\Gamma} dS \Big[  \vect{n}\cdot \kappa _1 \nabla \phi_1  +  (\frac{\partial f_s}{\partial \phi_{1 \Gamma}} -\kappa_{1 \Gamma} \Delta_{\Gamma} \phi_1) \Big] \frac{\partial \phi_{1\Gamma}}{\partial t} + \int_{\Gamma} dS \Big[  \vect{n}\cdot \kappa_2 \nabla \phi_2  +  (\frac{\partial f_s}{\partial \phi_{2 \Gamma}} -\kappa_{2 \Gamma} \Delta_{\Gamma} \phi_2) \Big] \frac{\partial \phi_{2\Gamma} }{\partial t} \nonumber\\
% &= \int_{\Omega} dx \Big[ \frac{\mu_1}{\nu}(\nabla\cdot M_1(\phi_1, \phi_2)  \nabla \mu_1 + \nabla\cdot M_{12}(\phi_1, \phi_2)  \nabla \mu_2) + \frac{\mu_2}{\nu}(\nabla\cdot M_2(\phi_1, \phi_2)  \nabla \mu_2 + \nabla\cdot M_{12}(\phi_1, \phi_2)  \nabla \mu_1) \Big]\nonumber\\
% & \quad + \int_{\Gamma} dS \Big[ F_1 (-\Gamma_1 F_1 - \Gamma_{12} F_2) + F_2 (-\Gamma_{12} F_1 - \Gamma_2 F_2) \Big] \nonumber\\
&= -\int_{\Omega} dx \frac{1}{\nu} \Big[ M_{1}(\phi_1, \phi_2) |\nabla\mu_1|^2 + 2M_{12}(\phi_1, \phi_2)\nabla\mu_1\cdot\nabla\mu_2 + M_{2}(\phi_1, \phi_2) |\nabla\mu_2|^2 \Big] \nonumber\\
& \quad - \int_{\Gamma} dS \Big[ \Gamma_1{{\mu_{1\Gamma}}}^2 + 2\Gamma_{12}{\mu_{1\Gamma}\mu_{2\Gamma}} + \Gamma_2 {{\mu_{2\Gamma}}}^2 \Big].
\end{align}
Since mobility matrices ${\left[ \begin{array}{cc}
M_1 & M_{12}\\
M_{12} & M_2\\
\end{array}\right ]}$, ${\left[ \begin{array}{cc}
\Gamma_1 & \Gamma_{12}\\
\Gamma_{12} & \Gamma_2\\
\end{array}\right ]}$
are positive definite, $\partial_t E \leq 0$ holds.

\subsection{Non-dimensionalization}
We set the characteristic length scale as the molecule length $l_0 = \nu^{1/3}$, characteristic time scale $t_0 = \nu^{2/3}/D_1$ such that $\tilde{D}_1 = D_1 t_0/l_0^2 = 1$.
Rescaling:
$\tilde{\vect{x}} = \vect{x}/  l_0$ and $\tilde{t} = t /t_0 $, our model has the following nondimensional parameters:
\begin{align}
\tilde{D}_{2} =\frac{D_2}{D_1} \, , \quad
\tilde{D}_{12} =\frac{D_{12}}{D_1} \, , \quad \tilde{\Gamma_i} = \Gamma_i \frac{k_{\text{B}}T}{\nu_{s}} t_0 \, , \quad
  {\bar{ \kappa}_i} =   {\kappa}_i  \frac{\nu_{s}}{k_{\text{B}}T}\frac{1}{l_0}= \frac{\nu^{1/3}}{l_0} \, ,\\
   \tilde{f_s} = f_s \frac{\nu_{s}}{k_{\text{B}}T} \, , \quad
 { \tilde{ \kappa }_{i\Gamma} =  \frac{1}{l_0^{2/3}} \kappa_{i\Gamma}  \frac{\nu_{s}}{k_{\text{B}}T} \, , \quad   \tilde{ \kappa }_i = \frac{1}{l_0^{2/3}} \kappa_{i}  \frac{\nu}{k_{\text{B}}T} }\,,  i = 1,2  .
\end{align}

For brevity, we skip the tildes in the following. The dimensionless equations governing the kinetics of the system are given as:
{
\begin{subequations}
\begin{align} \label{eq:gov_eqn_dim_less}
    \partial_t \phi_1 &= \nabla  \phi_1 (1-\phi_1-\phi_2) \cdot  \nabla {\mu}_1 + \nabla D_{12} \phi_1 \phi_2 (1-\phi_1-\phi_2) \cdot  \nabla {\mu}_2 , \\
    \partial_t \phi_2 &= \nabla  D_{12}  \phi_1 \phi_2 (1-\phi_1-\phi_2)  \cdot  \nabla {\mu}_1 +  \nabla  D_2  \phi_2 (1-\phi_1-\phi_2)  \cdot \nabla {\mu}_2,
    \end{align}
\end{subequations}}
with dimensionless boundary conditions are written as: 
\begin{subequations}
\begin{align}
\partial_t \phi_{1 \Gamma} &= -\Gamma_1 \Big[\vect{n} \cdot \bar{\kappa}_1 \nabla \phi_1 - \kappa_{1 \Gamma} \Delta_{\Gamma} \phi_{1 \Gamma} + \frac{\partial f_s}{\partial \phi_{1 \Gamma}} \Big] - \Gamma_{12} \Big[\vect{n} \cdot \bar{\kappa}_2 \nabla \phi_2 - \kappa_{2 \Gamma} \Delta_{\Gamma} \phi_{2 \Gamma} + \frac{\partial f_s}{\partial \phi_{2 \Gamma}} \Big]  , \\
   \partial_t \phi_{2 \Gamma} &= -\Gamma_2 \Big[\vect{n} \cdot \bar{\kappa}_2 \nabla \phi_2 - \kappa_{2 \Gamma} \Delta_{\Gamma} \phi_{2 \Gamma} + \frac{\partial f_s}{\partial \phi_{2 \Gamma}}\Big] - \Gamma_{12} \Big[\vect{n} \cdot \bar{\kappa}_1 \nabla \phi_1 - \kappa_{1 \Gamma} \Delta_{\Gamma} \phi_{1 \Gamma} + \frac{\partial f_s}{\partial \phi_{1 \Gamma}} \Big],\\
\partial_{\vect{n}} \mu_1|_{\Gamma} & = 0, \\
\partial_{\vect{n}} \mu_2|_{\Gamma} & = 0.
\end{align}
\end{subequations}
In the next section, we will develop the numerical scheme based on the Invariant Energy Quadratization method\cite{Yang_Zhao_Wang_ShenM3AS2017,Yang-EQ-PFC,ZYLW_SISC_2016,Jia_XF_YZ_XP_XG_Jun_Q2018}.
% ,Yang-EQ-PFC,ZYLW_SISC_2016,Jia_XF_YZ_XP_XG_Jun_Q2018}. 

\section{Numerical scheme}\label{sec:numericalscheme}
We first reformulate the mathematical model into an equivalent formation. Then we discretize the reformulated system in time and space, respectively. In this work, we consider the 2D domain for simplicity.
\subsection{Mathematical reformulation}
We reformulate the total free energy in a quadratic form
\begin{align}
    q_1 = \sqrt{ \frac{\phi_1}{n_1} \ln \phi_1 + \frac{\phi_2}{n_2} \ln \phi_2 + (1-\phi_1 - \phi_2) \ln(1-\phi_1 - \phi_2) + \chi_{12}\phi_1\phi_2 + \chi_{13} \phi_1\phi_3 + \chi_{23} \phi_2\phi_3 + C}, 
\end{align} 
where $C$ is a constant such that the value under the square root is positive, $\phi_3 = 1-\phi_1-\phi_2$.
Then, the total energy becomes
\begin{align}
    E = E_{\text{surf}} + E_{\text{bulk}} =  E_{\text{surf}} + \int_{\Omega} \Big[ q_1^2 + \omega_1 \phi_1 + \omega_2 \phi_2 + \frac{\kappa_1}{2}|\nabla \phi_1|^2 + \frac{\kappa_2}{2}|\nabla \phi_2|^2  \Big]d\vect{x}.
\end{align}
the chemical potentials become
\begin{subequations}
\begin{align}\label{eq:gov_eqn_reformulated}
    \mu_1 &= 2 q_1 \frac{\partial q_1}{\partial \phi_1} 
    + \omega_1 - \kappa_1 \Delta \phi_1 ,\\
    \mu_2 &= 2 q_1 \frac{\partial q_1}{\partial \phi_2} + \omega_2- \kappa_2 \Delta \phi_2.
\end{align}
\end{subequations}
The reformulated governing equations of the system are as follows.
\begin{subequations}
\begin{align}
    \partial_t \phi_1 &= \nabla  M_1(\phi_1, \phi_2) {\cdot} \nabla \mu_1 + \nabla  M_{12}(\phi_1, \phi_2) {\cdot} \nabla \mu_2 , \\
    \partial_t \phi_2 &= \nabla  M_2(\phi_1, \phi_2) {\cdot} \nabla \mu_2 + \nabla  M_{12}(\phi_1, \phi_2) {\cdot} \nabla \mu_1,\\
   \partial_t q_1  &= \frac{\partial q_1}{\partial \phi_1} \partial_t \phi_1 + \frac{\partial q_1}{\partial \phi_2} \partial_t \phi_2. 
\end{align}
\end{subequations}
with boundary conditions:
\begin{subequations}
\begin{align}
\partial_t \phi_{1 \Gamma} &= -\Gamma_1 \Big[\vect{n} \cdot \kappa_1 \nabla \phi_1 - \kappa_{1 \Gamma} \Delta_{\Gamma} \phi_{1 \Gamma} + \frac{\partial f_s}{\partial \phi_{1 \Gamma}} \Big] - \Gamma_{12} \Big[\vect{n} \cdot \kappa_2 \nabla \phi_2 - \kappa_{1 \Gamma} \Delta_{\Gamma} \phi_{2 \Gamma} + \frac{\partial f_s}{\partial \phi_{2 \Gamma}} \Big]  , \\
   \partial_t \phi_{2 \Gamma} &= -\Gamma_2 \Big[\vect{n} \cdot \kappa_2 \nabla \phi_2 - \kappa_{1 \Gamma} \Delta_{\Gamma} \phi_{2 \Gamma} + \frac{\partial f_s}{\partial \phi_{2 \Gamma}}\Big] - \Gamma_{12} \Big[\vect{n} \cdot \kappa_1 \nabla \phi_1 - \kappa_{1 \Gamma} \Delta_{\Gamma} \phi_{1 \Gamma} + \frac{\partial f_s}{\partial \phi_{1 \Gamma}} \Big],\\
   \partial_n \mu_1|_{\Gamma} &= 0,\\
   \partial_n \mu_2|_{\Gamma} &= 0.
\end{align}
\end{subequations}
The reformulated system has the following properties.
\begin{theorem}\label{eq:theorem_total_mass}
The total mass of each component in the domain $\Omega$ is conserved, i.e. 
\begin{align}
    \int_{\Omega} \phi_i(x,y,t) d\vect{x} =  \int_{\Omega} \phi_i(x,y,0) d\vect{x}, \quad i=1,2.
\end{align}

\end{theorem}

\begin{theorem}\label{eq:theorem_energy}
The total free energy is dissipative, i.e.
\begin{align}
    \partial_t E = \partial_t (E_{\text{\rm surf}} + E_{\text{\rm bulk}}) \leq  0.
\end{align}
\end{theorem}

% \begin{enumerate}
%     \item total mass conservation
%     \item total energy decreases with respect to time.
% \end{enumerate}
We skip the proof here. 
Next, we will discretize the reformulated governing equation in both the time and space directions.

\subsection{Semi-discrete scheme in time}
We use the Crank-Nicolson method in the time direction. $\Delta {t}$ is the time step. $(\cdot)^n$ represents the solution at {$n^{th}$} time step, i.e. ${t_n} = n\Delta t$. We denote
\begin{align}
    \partial_t^{n+1} (\cdot) = \frac{(\cdot)^{n+1}-(\cdot)^n}{\Delta t}, \quad 
    (\cdot)^{n+1/2} = \frac{(\cdot)^{n} + (\cdot)^{n+1}}{2}, \quad   \bar{(\cdot)}^{n+1/2} = \frac{3(\cdot)^{n} - (\cdot)^{n-1}}{2}.
\end{align}
The governing equations of the system are as follows:
\begin{subequations}
\begin{align}
    \partial_t^{n+1} \phi_1 &= \nabla M_1(\bar{\phi_1}^{n+1/2}, \bar{\phi_2}^{n+1/2}) {\cdot} \nabla \mu_1^{n+1/2} + \nabla M_{12}(\bar{\phi_1}^{n+1/2}, \bar{\phi_2}^{n+1/2}) {\cdot} \nabla \mu_2^{n+1/2} , \\
    \partial_t^{n+1} \phi_2 &= \nabla M_2(\bar{\phi_1}^{n+1/2}, \bar{\phi_2}^{n+1/2}) {\cdot} \nabla \mu_2^{n+1/2} + \nabla M_{12}(\bar{\phi_1}^{n+1/2}, \bar{\phi_2}^{n+1/2}) {\cdot} \nabla \mu_1^{n+1/2},\\
   \partial_t^{n+1} q_1 &= \bar{\frac{\partial q_1}{\partial \phi_1}}^{n+1/2} \partial_t^{n+1} \phi_1 +\bar{\frac{\partial q_1}{\partial \phi_2}}^{n+1/2} \partial_t^{n+1} \phi_2 .
\end{align}
\end{subequations}
with boundary conditions:
\begin{subequations}
\begin{align}  %\label{eq:bc_discrete}
\partial_t^{n+1} \phi_{1 \Gamma} &= -\Gamma_1 \Big[\vect{n} \cdot \bar{\kappa}_1 \nabla \phi_1^{n+1/2} - \kappa_{1 \Gamma} \Delta_{\Gamma} \phi_{1 \Gamma}^{n+1/2} + \frac{\partial f_s}{\partial \phi_{1 \Gamma}}^{n+1/2} \Big] \\ \nonumber
   & \quad  - \Gamma_{12} \Big[\vect{n} \cdot \bar{\kappa}_2 \nabla \phi_2^{n+1/2} - {\kappa}_{1 \Gamma} \Delta_{\Gamma} \phi_{2 \Gamma}^{n+1/2} + \frac{\partial f_s}{\partial \phi_{2 \Gamma}}^{n+1/2} \Big]  , \\
   \partial_t^{n+1} \phi_{2 \Gamma} &= -\Gamma_2 \Big[\vect{n} \cdot \bar{\kappa}_2 \nabla \phi_2^{n+1/2} - \kappa_{1 \Gamma} \Delta_{\Gamma} \phi_{2 \Gamma}^{n+1/2} + \frac{\partial f_s}{\partial \phi_{2 \Gamma}}^{n+1/2}\Big] \\ \nonumber 
   & \quad - \Gamma_{12} \Big[\vect{n} \cdot \bar{\kappa}_1 \nabla \phi_1^{n+1/2} - \kappa_{1 \Gamma} \Delta_{\Gamma} \phi_{1 \Gamma}^{n+1/2} + \frac{\partial f_s}{\partial \phi_{1 \Gamma}}^{n+1/2} \Big] ,\\
\partial_{\vect{n}} \mu_1^{n+1}|_{\Gamma} & = 0, \\
\partial_{\vect{n}} \mu_2^{n+1}|_{\Gamma} & = 0.
\end{align}
\end{subequations}
where
\begin{subequations}
\begin{align}
    \mu_1^{n+1/2} &= 2 q_1^{n+1/2} \bar{\frac{\partial q_1}{\partial \phi_1}}^{n+1/2}
    + \omega_1  - {\kappa}_1 \Delta \phi_1^{n+1/2} ,\\ 
    \mu_2^{n+1/2} &= 2 q_1^{n+1/2} \bar{\frac{\partial q_1}{\partial \phi_2}}^{n+1/2} + \omega_2 - {\kappa}_{2} \Delta \phi_2^{n+1/2}. 
\end{align}
\end{subequations}

\subsection{Fully discrete scheme}

\begin{figure}
		\scalebox{1}{
			\begin{tikzpicture} [line width=0.8, scale=1] 
				\clip (4,-2.5) rectangle (16,5.5);
				
				\draw [ line width=0.8] (10,0.5)--(10,2);
				
				%\draw[gray,fill=blue!30, very thin] (11.5,2.8)--(10.0,2.0)--(8.7,0.5)--(8.5,0.28)--(7.9, -1.0)--(11.5,-1.0)--(11.5,2.8)--cycle;

				\filldraw [black] (7.75, -0.25) circle (2pt);
				\filldraw [black] (7.75,1.25) circle (2pt);
				\filldraw [black] (7.75, 2.75) circle (2pt); 
				\filldraw [black](9.25,-0.25) circle (2pt);
				\filldraw [black] (9.25,1.25) circle (2pt);
				%node[anchor=north] {\small$\mathcal{C}_{i,j}$};
				\filldraw [black] (9.25,2.75) circle (2pt);
				\filldraw [black] (10.75,-0.25) circle (2pt);
				\filldraw [black](10.75,1.25) circle (2pt);
				\filldraw [black] (10.75,2.75) circle (2pt);
				
				% solid lines 
				\draw[blue, line width=0.8] (7.0,-1.0) -- (11.5,-1.0) ;
				\draw[black, line width=0.8] (7.0,0.5) -- (11.5,0.5) ;
				\draw[black, line width=0.8] (7.0,2.0) -- (11.5,2.0) ;
				\draw[blue, line width=0.8] (7.0,3.5) -- (11.5,3.5) ;
				
				\draw[blue, line width=0.8] (7.0,-1.0) -- (7.0,3.5) ;
				\draw[black, line width=0.8] (8.5,-1.0) -- (8.5,3.5) ;
				\draw[black, line width=0.8] (10,-1.0) -- (10,3.5) ;
				\draw[blue, line width=0.8] (11.5,-1.0) -- (11.5,3.5) ;
				
				%dashed line 
				%right
				\draw[dashed, gray, line width=0.8] (13,-1.0) -- (13,3.5);
				\draw[dashed,gray,line width=0.8](11.5,-1.0)--(13,-1.0);
				\draw[dashed,gray,line width=0.8](11.5,0.5)--(13,0.5);
				\draw[dashed,gray,line width=0.8](11.5,2.0)--(13,2.0);
				\draw[dashed,gray,line width=0.8](11.5,3.5)--(13,3.5);
				
				\filldraw [color=black, fill=white] (12.25, -0.25) circle (2pt);
				\filldraw [color=black, fill=white] (12.25,1.25) circle (2pt); 
				\filldraw [color=black, fill=white] (12.25,2.75) circle (2pt);

				%left
				\draw[dashed, gray, line width=0.8] (5.5,-1.0) -- (5.5,3.5);
				\draw[dashed,gray,line width=0.8](5.5,-1.0)--(7.0,-1.0);
				\draw[dashed,gray,line width=0.8](5.5,0.5)--(7.0,0.5);
				\draw[dashed,gray,line width=0.8](5.5,2.0)--(7.0,2.0);
				\draw[dashed,gray,line width=0.8](5.5,3.5)--(7.0,3.5);
				
				\filldraw [color=black, fill=white] (6.25, -0.25) circle (2pt);
				\filldraw [color=black, fill=white] (6.25,1.25) circle (2pt); 
				\filldraw [color=black, fill=white] (6.25,2.75) circle (2pt);

				%top
				\draw[dashed, gray, line width=0.8] (7.0,5) -- (11.5,5);
				\draw[dashed,gray,line width=0.8](7.0,3.5)--(7.0,5);
				\draw[dashed,gray,line width=0.8](8.5,3.5)--(8.5,5);
				\draw[dashed,gray,line width=0.8](10,3.5)--(10,5);
				\draw[dashed,gray,line width=0.8](11.5,3.5)--(11.5,5);
				
				\filldraw [color=black, fill=white] (7.75, 4.25) circle (2pt);
				\filldraw [color=black, fill=white] (9.25,4.25) circle (2pt); 
				\filldraw [color=black, fill=white] (10.75,4.25) circle (2pt);

				%bottom 
				\draw[dashed, gray, line width=0.8] (7.0,-2.5) -- (11.5,-2.5);
				\draw[dashed,gray,line width=0.8](7.0,-2.5)--(7.0,-1.0);
				\draw[dashed,gray,line width=0.8](8.5,-2.5)--(8.5,-1.0);
				\draw[dashed,gray,line width=0.8](10,-2.5)--(10,-1.0);
				\draw[dashed,gray,line width=0.8](11.5,-2.5)--(11.5,-1.0);
				
				\filldraw [color=black, fill=white] (7.75, -1.75) circle (2pt);
				\filldraw [color=black, fill=white] (9.25,-1.75) circle (2pt); 
				\filldraw [color=black, fill=white] (10.75,-1.75) circle (2pt);
				
				\filldraw [black] (13.6,5) circle (2pt)node[anchor=west] {\textcolor{black}{\small{ \rm Interior point }}};
				
				\filldraw [color=black, fill=white] (13.6,4.3) circle (2pt)node[anchor=west] {\small{ \rm Ghost point }};
    
                \draw (13.6,3.6) node[cross=3pt,black]{}node[anchor=west] {\small{ \rm Edge point }};

                %edge points
                %vertical lines
                \draw (7,2.75) node[cross=3pt,black] {};
                 \draw (7,1.25) node[cross=3pt,black] {};
                 \draw (7,-0.25) node[cross=3pt,black] {};

                 \draw (8.5,2.75) node[cross=3pt,black] {};
                 \draw (8.5,1.25) node[cross=3pt,black] {};
                 \draw (8.5,-0.25) node[cross=3pt,black] {};

                 \draw (10,2.75) node[cross=3pt,black] {};
                 \draw (10,1.25) node[cross=3pt,black] {};
                 \draw (10,-0.25) node[cross=3pt,black] {};

				\draw (11.5,2.75) node[cross=3pt,black] {};
                 \draw (11.5,1.25) node[cross=3pt,black] {};
                 \draw (11.5,-0.25) node[cross=3pt,black] {};

                  % Horizontal lines

                  \draw (7.75,3.5) node[cross=3pt,black] {};
                 \draw (9.25,3.5) node[cross=3pt,black] {};
                 \draw (10.75,3.5) node[cross=3pt,black] {};

                 \draw (7.75,2) node[cross=3pt,black] {};
                 \draw (9.25,2) node[cross=3pt,black] {};
                 \draw (10.75,2) node[cross=3pt,black] {};

                 \draw (7.75,0.5) node[cross=3pt,black] {};
                 \draw (9.25,0.5) node[cross=3pt,black] {};
                 \draw (10.75,0.5) node[cross=3pt,black] {};

                \draw (7.75,-1) node[cross=3pt,black] {};
                 \draw (9.25,-1) node[cross=3pt,black] {};
                 \draw (10.75,-1) node[cross=3pt,black] {};

				% x axis and y axis 
					\draw[->,line width=1](5,-2.5)--(5,-1.0)node[anchor=north,xshift=-6pt, yshift=-8pt] {\normalsize{$y$}};
						\draw [->,line width=1] (4.5,-2.0)--(6.0,-2.0)node[anchor=east,xshift=-6pt, yshift=-6pt] {\normalsize{$x$}};
				%\draw [-{Stealth[length=3mm, width=2mm]}] (4.5,-2.0)--(6.0,-2.0)node[anchor=east,xshift=-6pt, yshift=-6pt] {\normalsize{$x$}};
				
					%\draw [->new, arrowhead=0.25in, line width=4pt](4.5,-2.0)--(6.0,-2.0);

			\end{tikzpicture}
		    }
			\caption{Sketch of the staggered mesh in two dimensional space. The blue line is the boundary $\Gamma$ of the 2D domain $\Omega = [0, L]\times[0,L]$ in $x-y$ plane. {The black solid points are the center points inside the bulk, the black empty points are the ghost points adjacent to the boundary $\Gamma$, and the crosses denote the edge points.} }
   % \SL{line width changed,  pointer size changed. solid gray points, size of text has been adjusted}
			% \com{Shuang, can you make the lines thicker, and pointer larger? Can you add the edge points in the plots using gray color?  if the text x and y can be larger, that would be great.  }

   \label{fig:mesh}
	\end{figure}
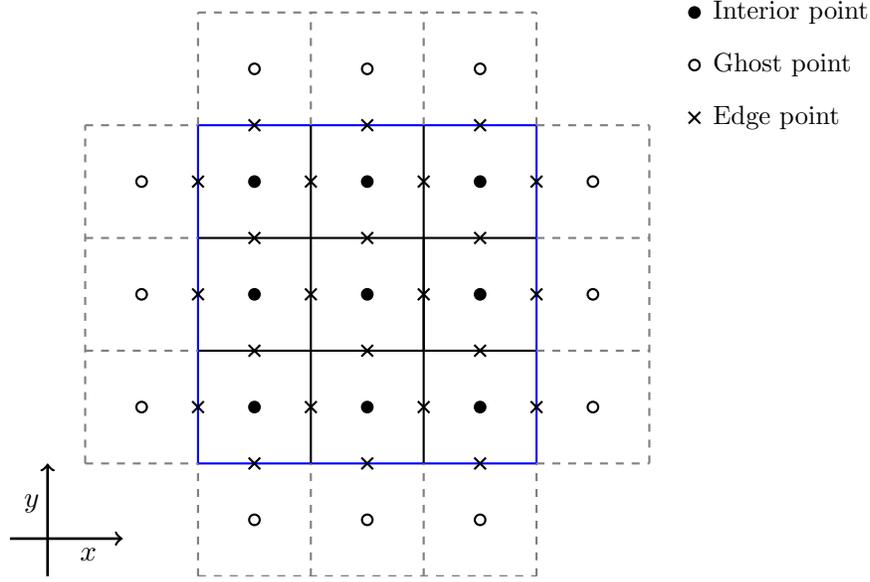

We use central finite difference method such that there is second order accuracy in the space. In specific,
We use uniform mesh in two-dimensional space $[0, L] \times [0, L]$ (see Fig.~\ref{fig:mesh}). {In the $x$-direction, the domain is divided into $N_x$ equal-sized subintervals. Similarly, in the $y$-direction, the domain is divided into $N_y$ subintervals, each of equal size. We discretize the scalar functions such as $\phi_i$ and $\Delta \phi_i$ values at the center point $(x_i, y_j)$, where \begin{align}
x_i &= (i - \frac{1}{2})\frac{L}{N_x}, \qquad i = 0, 1, \cdots, N_x+1,\\ \nonumber
y_j &= (j - \frac{1}{2})\frac{L}{N_y}, \qquad j = 0, 1, \cdots, N_y+1, 
\end{align} while discretize the vector functions, e.g. $\nabla \phi_i$ at the edge points $(x_{i+\frac{1}{2}}, y_{j})$ or $(x_{i}, y_{j+\frac{1}{2}})$\cite{Zhao_W2018_2}, where \begin{align}
x_{i+\frac{1}{2}} &= i\frac{L}{N_x}, \qquad i = 0, 1, \cdots, N_x,\\ \nonumber
y_{j+\frac{1}{2}} &= j\frac{L}{N_y}, \qquad j = 0, 1, \cdots, N_y, 
\end{align} }  
At the boundary $\Gamma$, we use the average value at adjacent discrete center points. e.g.
{
\begin{align}
    \phi_{1\Gamma}^n|_{\frac{1}{2},j} = \frac{\phi_1^n|_{0,j}+\phi_1^n|_{1,j}}{2}, \qquad \phi_{1\Gamma}^n|_{N_x+\frac{1}{2},j} = \frac{\phi_1^n|_{N_x,j}+\phi_1^n|_{N_x+1,j}}{2}.
\end{align}}
$j = 1, \cdots N_y$, is the index in $y$-direction.

{
We denote the edge-to-center average and difference operator as $a_x$, $d_x$ in {$x$-direction}, $a_y$, $d_y$ in {$y$-direction}, which are defined as follows
\begin{align}
  a_xu_{i,j}:=\frac{1}{2}(u_{i+\frac{1}{2},j}+u_{i-\frac{1}{2},j}),\quad &d_xu_{i,j}:=\frac{1}{h_x}(u_{i+\frac{1}{2},j}-u_{i-\frac{1}{2},j}), \\
  a_yv_{i,j}:=\frac{1}{2}(v_{i,j+\frac{1}{2}}+v_{i,j-\frac{1}{2}}),\quad &d_yu_{i,j}:=\frac{1}{h_y}(v_{i,j+\frac{1}{2}}-v_{i,j-\frac{1}{2}}).
  \end{align}

  We denote the center-to-edge average and difference operators by $A_x$, $D_x$ in $x$-direction. Analogously, the center-to-edge in $y$-direction average and difference operators are denoted as $A_y$, $D_y$. Their definitions are as follows.
  \begin{align}
   A_x\phi_{i+\frac{1}{2},j}:=\frac{1}{2}(\phi_{i+1,j}+\phi_{i,j}),\quad &D_x\phi_{i+\frac{1}{2},j}:=\frac{1}{h_x}(\phi_{i+1,j}-\phi_{i,j}),\\
  A_y\phi_{i,j+\frac{1}{2}}:=\frac{1}{2}(\phi_{i,j+1}+\phi_{i,j}),\quad &D_y\phi_{i,j+\frac{1}{2}}:=\frac{1}{h_y}(\phi_{i,j+1}-\phi_{i,j}).
\end{align}

  The standard 2D discrete Laplace operator is defined as ${\Delta}_h$ in the bulk and ${\Delta}_{\Gamma h}$ on the surface. 
  {
\begin{equation}
    {\Delta}_h\phi :=d_x(D_x\phi)+d_y(D_y\phi),\quad  {\Delta}_{\Gamma h }\phi_{\Gamma} :=d_y(D_y\phi_{\Gamma}), \quad \text{or}\quad  {\Delta}_{\Gamma h }\phi_{\Gamma} :=d_x(D_x\phi_{\Gamma}).
\end{equation}}

The discrete norm in bulk and surface are defined as follows:
\begin{align}
 ||\phi ||_c = h_x h_y \sum_{i=1}^{N_x} \sum_{j=1}^{N_y} \phi_{i,j},  \qquad ||\nabla \phi ||_v = h_x h_y \sum_{i=1}^{N_x} \sum_{j=1}^{N_y} \Big[a_xD_x \phi_{i+1/2,j} + a_yD_y \phi_{i,j+1/2}\Big],
 \end{align}
 
\begin{align}
 ||\phi ||_{c,\Gamma} =  h_y \sum_{j=1}^{N_y} \phi_{j\Gamma} ,  \qquad ||\nabla \phi ||_{v,\Gamma} = h_y  \sum_{j=1}^{N_y} \Big[ a_yD_y \phi_{i,j+1/2}\Big].
\end{align}

\textbf{The fully discrete numerical scheme:}
\begin{align}
    \Big \{\partial_t^{n+1} \phi_1 &= d_x A_x(M_1(\bar{\phi_1}^{n+1/2}, \bar{\phi_2}^{n+1/2})  D_x \mu_1^{n+1/2} + d_y A_y(M_1(\bar{\phi_1}^{n+1/2}, \bar{\phi_2}^{n+1/2}))  D_y \mu_1^{n+1/2} \nonumber\\
    & + d_x A_x(M_{12}(\bar{\phi_1}^{n+1/2}, \bar{\phi_2}^{n+1/2})  D_x \mu_2^{n+1/2} + d_y A_y(M_{12}(\bar{\phi_1}^{n+1/2}, \bar{\phi_2}^{n+1/2}))  D_y \mu_2^{n+1/2} \Big \}\Big|_{i,j},  \\
    \Big \{\partial_t^{n+1} \phi_2 &= d_x A_x(M_2(\bar{\phi_1}^{n+1/2}, \bar{\phi_2}^{n+1/2})  D_x \mu_2^{n+1/2} + d_y A_y(M_2(\bar{\phi_1}^{n+1/2}, \bar{\phi_2}^{n+1/2}))  D_y \mu_2^{n+1/2}\nonumber \\
    & + d_x A_x(M_{12}(\bar{\phi_1}^{n+1/2}, \bar{\phi_2}^{n+1/2})  D_x \mu_1^{n+1/2} + d_y A_y(M_{12}(\bar{\phi_1}^{n+1/2}, \bar{\phi_2}^{n+1/2}))  D_y \mu_1^{n+1/2} \Big \}\Big|_{i,j},\\
   \Big \{\partial_t^{n+1} q_1 &= \bar{\frac{\partial q_1}{\partial \phi_1}}^{n+1/2} \partial_t^{n+1} \phi_1 +\bar{\frac{\partial q_1}{\partial \phi_2}}^{n+1/2} \partial_t^{n+1} \phi_2\Big \}\Big|_{i,j} \, .
\end{align}
where $i = 1, \cdots N_x$, $j = 1, \cdots N_y$.
   % % \SL{$\Big \{\partial_t^{n+1} q_1 &= \bar{\frac{\partial q_1}{\partial \phi_1}}|_{i,j}^{n+1/2} \partial_t^{n+1} \phi_1|_{i,j} +\bar{\frac{\partial q_1}{\partial \phi_2}}|_{i,j}^{n+1/2} \partial_t^{n+1} \phi_2_{i,j}\Big \}\Big|_{i,j} \,. $}
   % \SL{$(i,j)$ is repeated, choose one type}
with boundary conditions:
\begin{subequations}\label{eq:bc_discrete}
    \begin{align}
\Big \{ \partial_t^{n+1} \phi_{1 \Gamma} &= -\Gamma_1 \Big[\vect{n} \cdot \bar{\kappa}_1 d_x \phi_1^{n+1/2} - \kappa_{1 \Gamma} \Delta_{\Gamma h} \phi_{1 \Gamma}^{n+1/2} + \frac{\partial f_s}{\partial \phi_{1 \Gamma}}^{n+1/2} \Big] \\ \nonumber 
   & \quad  - \Gamma_{12} \Big[\vect{n} \cdot \bar{\kappa}_2 d_x \phi_2^{n+1/2} - {\kappa}_{2 \Gamma} \Delta_{\Gamma h} \phi_{2 \Gamma}^{n+1/2} + \frac{\partial f_s}{\partial \phi_{2 \Gamma}}^{n+1/2} \Big]  \Big \}\Big|_{\frac{1}{2}\text{or} ({N_x}+\frac{1}{2})  ,j}, \\
  \Big \{ \partial_t^{n+1} \phi_{2 \Gamma} &= -\Gamma_2 \Big[\vect{n} \cdot \bar{\kappa}_2 d_x \phi_2^{n+1/2} - \kappa_{2 \Gamma} \Delta_{\Gamma h} \phi_{2 \Gamma}^{n+1/2} + \frac{\partial f_s}{\partial \phi_{2 \Gamma}}^{n+1/2}\Big] \\ \nonumber 
   & \quad - \Gamma_{12} \Big[\vect{n} \cdot \bar{\kappa}_1 d_x \phi_1^{n+1/2} - \kappa_{1 \Gamma} \Delta_{\Gamma h} \phi_{1 \Gamma}^{n+1/2} + \frac{\partial f_s}{\partial \phi_{1 \Gamma}}^{n+1/2} \Big] \Big \}\Big|_{\frac{1}{2} \text{or} ({N_x}+\frac{1}{2})  ,j},\\
 \mu_{{k}}|_{0,j} &= \mu_{{k}}|_{1,j}, \qquad
 \mu_{{k}}|_{{N_x},j} = \mu_{{k}}|_{{N_x}+1,j}, \qquad k = 1,2. \\
 \mu_{{k}}|_{i,0} &= \mu_{{k}}|_{i,1}, \qquad
 \mu_{{k}}|_{i, {N_y}} = \mu_{{k}}|_{i,{N_y}+1}, \qquad k = 1,2,
\end{align}
\end{subequations}where
\begin{subequations}\label{eq:discrete_mus}
    \begin{align} 
    \mu_1^{n+1/2}|_{i,j} &= 2 q_1|_{i,j}^{n+1/2} \bar{\frac{\partial q_1}{\partial \phi_1}}^{n+1/2}|_{i,j}
    + \omega_1  - {\kappa}_1 \Delta_h \phi_1^{n+1/2}|_{i,j} ,\\ 
    \mu_2^{n+1/2}|_{i,j} &= 2 q_1|_{i,j}^{n+1/2} \bar{\frac{\partial q_1}{\partial \phi_2}}^{n+1/2}|_{i,j} + \omega_2 - {\kappa}_{2} \Delta_h \phi_2^{n+1/2}|_{i,j}. 
\end{align}
\end{subequations}

The fully discrete numerical scheme fulfills the total mass conservation law and the total energy dissipation rate at the discrete level.

% \com{Update here to theorems}\SL{comments?}
\begin{theorem}\label{eq:total_mass_discrete}
\textbf{Mass conservation law at discrete level:}
Based on the discrete boundary conditions \eqref{eq:bc_discrete}
\begin{align}
   \sum_{i=1}^{N_x}\sum_{j=1}^{N_y} \frac{\phi_1^{n+1}|_{i,j} - \phi_1^n|_{i,j}}{\Delta t} = \sum_{i=1}^{N_x}\sum_{j=1}^{N_y} \Big [ \nabla \cdot M_1 \nabla \mu_1^{n+1/2}|_{i,j}  + \nabla \cdot M_{12} \nabla \mu_2^{n+1/2}|_{i,j} \Big] = 0. 
\end{align}
same as for the total mass of $\phi_2$.

\end{theorem}

\begin{theorem}\label{eq:total_energy_discrete}
\textbf{Unconditionally energy stability at discrete level:}
The discrete energy of the system is defined as:
%\begin{subequations} 
\begin{align}\label{eq:discrete_Energy}
E^{n+1} = E_{\text{\rm bulk}}^{n+1} + E_{\text{\rm surf}}^{n+1},
\end{align}
where
\begin{align}
E_{\text{\rm bulk}}^{n+1} &= ||  f_b(\phi_1, \phi_2)^{n+1}||_c  + \frac{\kappa_1}{2}||\nabla \phi_1^{n+1}||_v^2 + \frac{\kappa_2}{2}||\nabla \phi_2^{n+1}||_v^2 ,\\
E_{\text{\rm surf}}^{n+1} & = ||f_s(\phi_{1 \Gamma}, \phi_{2 \Gamma})^{n+1}||_{c,\Gamma} + \frac{\kappa_{1\Gamma}}{2}||\nabla \phi_{1 \Gamma}^{n+1}||_{v,\Gamma}^2 + \frac{\kappa_{2\Gamma}}{2}||\nabla \phi_{2 \Gamma}^{n+1}||_{v,\Gamma}^2.
\end{align}

%\end{subequations}

% \com{double check the range at edge points for norm calculation.}
% \com{rewrite the proof in format of norm and add lemmas}
The total energy at discrete level decreases with respect to time, i.e.
\begin{align}\label{eq:Energy_dissipation_discrete}
    E^{n+1} \leq E^n.
\end{align}
\end{theorem}
\textbf{proof:}
Using the definition of reformulated total energy at discrete level \eqref{eq:discrete_Energy} and the formula $(a^2 - b^2) = (a + b) (a -b)$ we obtain, 
% Combining lemmas \eqref{}, conservation law \eqref{} and boundary conditions \eqref{}, we have the discretized energy dissipation rate as follows
% % \com{Write down the proof steps one by one.} \SL{like follows?}
% \begin{align}
%     \frac{E^{n+1}-E^n}{\Delta t} =& \sum_{i=1}^N \Big[ f_b(\phi_1, \phi_2)^{n+1} \Big]  + \sum_{i=1/2}^{N+1/2} \Big [ \frac{\kappa_1}{2}|\nabla \phi_1^{n+1}|^2 + \frac{\kappa_2}{2}|\nabla \phi_2^{n+1}|^2 \Big] \nonumber\\
%     & +  2\Big[f_s(\phi_{1 \Gamma}, \phi_{2 \Gamma})^{n+1} + \frac{\kappa_{1\Gamma}}{2}|\nabla \phi_{1 \Gamma}^{n+1}|^2 + \frac{\kappa_{2\Gamma}}{2}|\nabla \phi_{2 \Gamma}^{n+1}|^2 \Big]\nonumber\\
%     & -\sum_{i=1}^N \Big[ f_b(\phi_1, \phi_2)^{n+1}  \Big]  + \sum_{i=1/2}^{N+1/2} \Big [ \frac{\kappa_1}{2}|\nabla \phi_1^{n+1}|^2 + \frac{\kappa_2}{2}|\nabla \phi_2^{n+1}|^2 \Big] \nonumber\\
%     &  -  2\Big[f_s(\phi_{1 \Gamma}, \phi_{2 \Gamma})^{n+1} + \frac{\kappa_{1\Gamma}}{2}|\nabla \phi_{1 \Gamma}^{n+1}|^2 + \frac{\kappa_{2\Gamma}}{2}|\nabla \phi_{2 \Gamma}^{n+1}|^2 \Big]\nonumber
%     \end{align}
    \begin{align}
    \frac{E^{n+1}-E^n}{\Delta t}  =&|| 2\cdot \frac{q_1^{n+1}-q_1^n}{\Delta t} \, \frac{q_1^{n+1} + q_1^n}{2} + \omega_1 \frac{\phi_1^{n+1}-\phi_1^n}{\Delta t} + \omega_2 \frac{\phi_2^{n+1}-\phi_2^n}{\Delta t}||_c \nonumber\\
    & +  \frac{\kappa_1}{2} ||2\cdot \frac{\nabla \phi_1^{n+1}-\nabla \phi_1^n}{\Delta t} \, \frac{\nabla \phi_1^{n+1} + \nabla \phi_1^n}{2}||_v + \frac{\kappa_2}{2} ||2\cdot \frac{\nabla \phi_2^{n+1}-\nabla \phi_2^n}{\Delta t} \, \frac{\nabla \phi_2^{n+1} + \nabla \phi_2^n}{2}||_v \nonumber\\
    & + \frac{\kappa_{1\Gamma}}{2} ||2\cdot \frac{\nabla_{||} \phi_{1\Gamma}^{n+1}-\nabla_{||} \phi_{1\Gamma}^n}{\Delta t} \, \frac{\nabla_{||} \phi_{1\Gamma}^{n+1} + \nabla_{||} \phi_{1\Gamma}^n}{2}||_{v,\Gamma} \nonumber \\
    & + \frac{\kappa_{2\Gamma}}{2} ||2\cdot \frac{\nabla_{||} \phi_{2\Gamma}^{n+1}-\nabla_{||} \phi_{2\Gamma}^n}{\Delta t} \, \frac{\nabla_{||} \phi_{2\Gamma}^{n+1} + \nabla_{||} \phi_{2\Gamma}^n}{2}||_{v,\Gamma} \nonumber\\
    & + h_1 ||\frac{\phi_1^{n+1}-\phi_1^n}{\Delta t} ||_c + g_1 ||2\cdot \frac{\phi_1^{n+1}-\phi_1^n}{\Delta t} \frac{\phi_1^{n+1} + \phi_1^n}{2}||_c \nonumber \\
    & + h_2||\frac{\phi_2^{n+1}-\phi_2^n}{\Delta t}||_c + g_2 ||\frac{\phi_2^{n+1}-\phi_2^n}{\Delta t} \frac{\phi_2^{n+1} + \phi_2^n}{2} ||_c\nonumber\\
    & + \gamma ||\frac{\phi_1^{n+1} + \phi_1^n}{2} \, \frac{\phi_2^{n+1} - \phi_2^n}{\Delta t}||_c + \gamma ||\frac{\phi_2^{n+1} + \phi_2^n}{2} \, \frac{\phi_1^{n+1} - \phi_1^n}{\Delta t}||_c.  
     \end{align}
     According to the discrete chemical potentials \eqref{eq:discrete_mus} and discrete boundary conditions \eqref{eq:bc_discrete} we have
     \begin{align}
     \frac{E^{n+1}-E^n}{\Delta t}  
      % &= \sum_{i=1}^N \Big[ \mu_1^{n+1/2} \, \frac{\phi_1^{n+1}-\phi_1^n}{\Delta t} + \mu_2^{n+1/2} \, \frac{\phi_2^{n+1}-\phi_2^n}{\Delta t} \Big ]+ \frac{\phi_{2\Gamma}^{n+1} - \phi_{2\Gamma}^{n}}{\Delta t} \, \frac{\nabla \phi_{2\Gamma}^{n+1} + \nabla \phi_{2\Gamma}^{n}}{2} \nonumber\\
      % & \quad + \frac{\phi_{1\Gamma}^{n+1} - \phi_{1\Gamma}^{n}}{\Delta t} \, \frac{\nabla \phi_{1\Gamma}^{n+1} + \nabla \phi_{1\Gamma}^{n}}{2} + \mu_{1\Gamma}^{n+1/2} \, \frac{\phi_{1 \Gamma}^{n+1}-\phi_{1 \Gamma}^n}{\Delta t}+ \mu_{2\Gamma}^{n+1/2} \, \frac{\phi_{2 \Gamma}^{n+1}-\phi_{2 \Gamma}^n}{\Delta t}\nonumber\\
      % & = - \sum_{i=1}^N \Big[ M_1 \nabla \mu_1^{n+1/2} \, \cdot  \nabla \mu_1^{n+1/2} + M_2 \nabla \mu_2^{n+1/2} \, \cdot \nabla \mu_2^{n+1/2}\Big ]\nonumber\\
      % & \quad + \frac{\phi_{1\Gamma}^{n+1} - \phi_{1\Gamma}^{n}}{\Delta t} \, \mu_{1\Gamma}^{n+1/2} + \frac{\phi_{2\Gamma}^{n+1} - \phi_{2\Gamma}^{n}}{\Delta t} \, \mu_{2\Gamma}^{n+1/2} \nonumber \\
      % & = - M_1 || \nabla \mu_1^{n+1/2} ||_{v}^2 - M_2 ||\nabla \mu_2^{n+1/2} ||_{v}^2   - \Gamma_{1} || \mu_{1\Gamma}^{n+1/2} ||_{v,\Gamma}^2 - \Gamma_{2} ||  \mu_{2\Gamma}^{n+1/2} ||_{v,\Gamma}^2 \leq 0.\nonumber
     & = ||\mu_1^{n+1/2}\cdot(\nabla  M_1 {\cdot} \nabla \mu_1^{n+1/2} + \nabla  M_{12} {\cdot} \nabla \mu_2^{n+1/2}) ||_{v} \nonumber\\
     & \quad + ||\mu_2^{n+1/2}\cdot(\nabla  M_2 {\cdot} \nabla \mu_2^{n+1/2} + \nabla  M_{12} {\cdot} \nabla \mu_1^{n+1/2}) ||_{v} \nonumber\\
     & \quad + ||\mu_{1\Gamma}^{n+1/2}\cdot(-\Gamma_{1} \mu_{1\Gamma}^{n+1/2} - \Gamma_{12} \mu_{2\Gamma}^{n+1/2})||_{v,\Gamma} \nonumber\\
     & \quad + ||\mu_{2\Gamma}^{n+1/2}\cdot(-\Gamma_{2} \mu_{2\Gamma}^{n+1/2} - \Gamma_{12} \mu_{1\Gamma}^{n+1/2})||_{v,\Gamma} \nonumber\\
     & = - || M_1 |\nabla \mu_1^{n+1/2}|^2 + 2M_{12}\nabla \mu_1^{n+1/2}\cdot \nabla \mu_2^{n+1/2} + M_2|\nabla \mu_2^{n+1/2}|^2 ||_{v} \nonumber\\
     & \quad - || \Gamma_{1} |\mu_{1\Gamma}^{n+1/2}|^2 + 2\Gamma_{12}\mu_{1\Gamma}^{n+1/2}\mu_{2\Gamma}^{n+1/2} + \Gamma_{2} |\mu_{2\Gamma}^{n+1/2}|^2 ||_{v,\Gamma}^2 \leq 0.
\end{align}
i.e. Eq.~\eqref{eq:Energy_dissipation_discrete} is obtained. 

\section{Physical phenomena investigation by numerical simulations }\label{sec:parameterstudy}

% The use of numerical simulations has several advantages, including the ability to study complex systems that may be difficult or impossible to study experimentally, the ability to vary parameters in a controlled manner, and the ability to study phenomena over a wide range of scales. In this study, the numerical simulations were conducted using a combination of the Invariant Energy Quadratization method, the Finite Difference Method in space, and the Crank-Nicolson scheme in time. The Invariant Energy Quadratization method \cite{} is a mathematical technique that involves reformulating the free energy of a system as a quadratic form using new auxiliary variables. This allows the system to be solved linearly and accurately. Once the mathematical model was reformulated using the Invariant Energy Quadratization method, the Finite Difference Method was used to approximate the derivatives in the reformulated system. The Crank-Nicolson scheme was then used to discretize the reformulated system in time.  

Taking advantages of numerical simulations on its ability to vary parameter in a controlled manner and over a wide range of scales, we investigate several physical phenomena in this section, including the wettability of multi-component droplets on solid surfaces, the effects of wall-mixture interactions on patterns in the bulk, and the role of cross-coupling relaxation rates in controlling kinetic processes in both the bulk and surface.

According to the Gibbs's rule, a ternary mixture system can have at most three coexisting phases in a system {(see Appendix~\ref{sec:appendix1} for details)}. In this section, we will evidence our points in both two-phase and three-phase coexisting scenarios. 

\subsection{Wettability, i.e. the contact angle is determined by the additive effects of wall-mixture interaction} 

The contact angle is a measure of the wetting behavior of a liquid on a solid surface. It is determined by the balance of forces between the solid phase, and other two liquid phases on the top of the solid surface. The wall-mixture interaction plays a significant role in determining the contact angle. 
% The strength of the interaction is influenced by various factors, such as the chemical and physical properties of the liquid and the solid surface, and the presence of any surfactants or additives. 
% The contact angle is the result of the additive effects of these forces, and it can be used to predict the wetting behavior of the liquid on the solid surface.
In the case of a multi-component liquid mixture, the contact angle is determined by the combined effects of the interactions between the individual component of the mixture and the solid surface. To grasp {these} additive effects, we define the surface free energy as the summation of {interactions} between individual component and the solid wall.

% Our aim in this section is to understand qualitatively
% how the contact angle $\theta$, depends on the surface free energy.

% We firstly investigate the effects in a two-phase system, and three-phase system later on. 
Specifically, the surface free energy reads
\begin{align}
     {f_s(\phi_{1\Gamma}, \phi_{2\Gamma)}} & = \frac{k_B T}{\nu_s} { \Big[ h_1 \phi_{1\Gamma} + g_1 \phi_{1\Gamma}^2 + h_2 \phi_{2\Gamma} + g_2 \phi_{2\Gamma}^2 + \gamma \phi_{1\Gamma} \phi_{2\Gamma} \Big]},
     % f_s(\phi_1, \phi_2) & = \frac{k_B T}{\nu_s} \Big[ h_1 \phi_1 + g_1 \phi_1^2 + h_2 \phi_2 + g_2 \phi_2^2 + \gamma \phi_1 \phi_2\Big],
\end{align}
Here, $h_i \phi_i$ depicts the interaction between solid wall and {i-th} component, $g_i \phi_i^2$ represents the interaction among {i-th} component molecules near by the solid surface. $\gamma \phi_1 \phi_2$ denotes the interaction between different kinds of molecules.

If all the {interactions} mentioned above are mutual, i.e. $h_i = g_i = \gamma = 0$, the contact angle of droplets on the surface is {$90^{\text{o}}$(see Fig.~\ref{fig:h12_phi_A}-e).} 
If the {interactions} between each component and solid wall are all attractive, i.e. $h_i < 0$, $g_i < 0$ or $\gamma < 0$, the contact angle will decrease until complete wetting (see Fig~\ref{fig:h12_phi_A},~\ref{fig:h12_phi_R},~\ref{fig:g12_phi_A},~\ref{fig:g12_phi_R}-g, and Fig.~\ref{fig:gamma_phi_a}-(a,b)), while repulsive {interactions} will lead to larger contact angle, i.e. $\theta > 90^{\text{o}}$(see Fig~\ref{fig:h12_phi_A},~\ref{fig:h12_phi_R},~\ref{fig:g12_phi_A},~\ref{fig:g12_phi_R}-c). If interactions between individual component and solid wall are different, e.g. the interaction between $\phi_1$ and solid wall is attractive, while the interaction between $\phi_2$ and the solid wall is repulsive, the effective interaction between droplet and solid wall is determined by the dominant component which has high {a higher} concentration in the droplet(see Fig~\ref{fig:h12_phi_A},~\ref{fig:h12_phi_R},~\ref{fig:g12_phi_A},~\ref{fig:g12_phi_R}-(a,i)). These observations can be verified in the three phases coexisting (see Appendix) wetting phenomena (see Fig.~\ref{fig:three_phases_g12_phi_A},~\ref{fig:three_phases_g12_phi_R}).

\subsection{Wall-mixture interaction changes the condensates patterns {far away from the surface}}

The influence of wall-mixture interactions {in} the formation of condensate patterns in systems exhibiting three-phase coexistence has yet to be extensively investigated. This study aims to bridge this gap by delving into the phase separation and condensates coarsening processes within such systems, paying {a} special attention to the varying types of wall-mixture interactions.

Our {numerical studies} revealed that interactions between the wall and the mixture exert a profound effect on the condensate patterns in the bulk, even {in} regions distant from the surface. We observed distinct variations in the condensate configurations contingent on the nature of the wall-mixture interactions. For instance, in cases where strong attractive or repulsive forces were at play, the resultant patterns manifested as strips running parallel to the solid wall, as evidenced by Fig.\ref{fig:PS_g12_phi_A} and Fig.\ref{fig:PS_g12_phi_R}-(a, b, c, d). Conversely, when the interactions were weak, the patterns that emerged were less structured and more sporadic, as can be seen in Fig.\ref{fig:PS_g12_phi_A} and Fig.\ref{fig:PS_g12_phi_R}-e. {These findings underscore the significance of wall-mixture interactions in determining the arrangement of condensates within the bulk.}

\subsection{Relaxation rate controls the kinetic process in the bulk and surface }

The relaxation rate serves as a pivotal factor influencing the kinetic processes occurring within the bulk and on the surface of a system. To elucidate the impact of kinetic rates {in} the spreading of droplets on a solid surface, we placed a droplet atop the surface and varied the kinetic rates, denoted as $\Gamma_i$, from $10^{-4}$ to 1. Our observations revealed that the surface relaxation rates play a decisive role in dictating the kinetics of droplet spreading. Specifically, a faster surface relaxation rate accelerates the spreading of the droplet, as depicted in Fig.~\ref{fig:spreading_panel1}.

The exploration of the effects of cross-coupling coefficients has been limited in {previous} research. In this study, we altered the cross-coupling {coefficient}, $M_{12}$, in the bulk to investigate {its} impact on the phase separation process therein. Our simulations indicate that changes in cross-coupling can indeed alter the kinetic processes (see Fig.\ref{fig:M12_phi_A} and Fig.\ref{fig:M12_phi_R} for the snapshots). {Total mass of $\phi_1$ and $\phi_2$ (see Fig.~\ref{fig:M12_Mass}) with different cross-coupling coefficient $M_{12}$ values are constants as proved by Theorem~\ref{eq:total_mass_discrete}. From the total energy profiles (see Fig.~\ref{fig:M12_Energy}), it is evident that variations in $M_{12}$ influence the total energy evolution during the initial stages of spontaneous phase separation.} Further, we adjusted the cross-coupling coefficient on the surface, $\Gamma_{12}$, to scrutinize its effects on the kinetics of droplet spreading on the solid surface. Our findings suggest that cross-coupling exerts a minimal effect on droplet spreading and coarsening subsequent to the phase separation process. 
% This can be attributed to the fact that the processes of droplet spreading and coarsening are predominantly influenced by the surface tension within the bulk. Thus, when the wall-mixture interaction and {diagonal surface relaxation rates $\Gamma_{i}, i=1,2,$} are held constant, variations in the surface cross-coupling relaxation rate {$\Gamma_{12}$} do not significantly impact the kinetics within the bulk.

% \SL{The role of relaxation rates and cross-coupling coefficients in affecting various processes like droplet spreading and phase separation is quite intriguing. 
% We first start with the experiment with varying surface relaxation rates and observing their impact on droplet spreading. It appears that faster surface relaxation rates accelerate the spreading of the droplet, suggesting a direct correlation between surface kinetics and the spreading process.
% Surprisingly, we explore into the effects of cross-coupling coefficients, cross-coupling exerts a minimal effect on droplet spreading and coarsening subsequent to the phase separation process. 
}

% Relaxation rate play a critical role in the kinetic process in both the bulk and surface of a system. To illustrate the influence of kinetic rates on the droplet spreading process on the solid surface, we put a droplet on the top of the surface and change the kinetic rates $\Gamma_i$ from $10^{-4}$ to 1. We observe that surface relaxation rates controls the droplet spreading kinetics. Faster relaxation in the surface can speed up the droplet spreading(see Fig.~\ref{fig:spreading_panel1}). 

% Effects of cross-coupling coefficients are rarely explored. Here, we change the cross-coupling coefficients $M_{12}$ in the bulk to explore its effects on the phase separation process in the bulk. Our simulation results show that cross-coupling change kinetic process (see Fig.~\ref{fig:M12_phi_A}, Fig.~\ref{fig:M12_phi_R}.)

% We further change the cross-coupling coefficient in the surface $\Gamma_{12}$ to investigate the cross-coupling effects on the droplet spreading kinetics on the top of the solid surface.
% We find that the cross-coupling has tiny effects on the droplet spreading and coarsening following the phase separation process. The reason is that the droplet spreading and coarsening process are mainly involved with the surface tension in the bulk. If the wall-mixture interaction and the diagonal relaxation rates are fixed, the cross-coupling relaxation rate will not influence the kinetics in the bulk a lot.

\begin{figure}
\centering
{\includegraphics[width=1\textwidth]{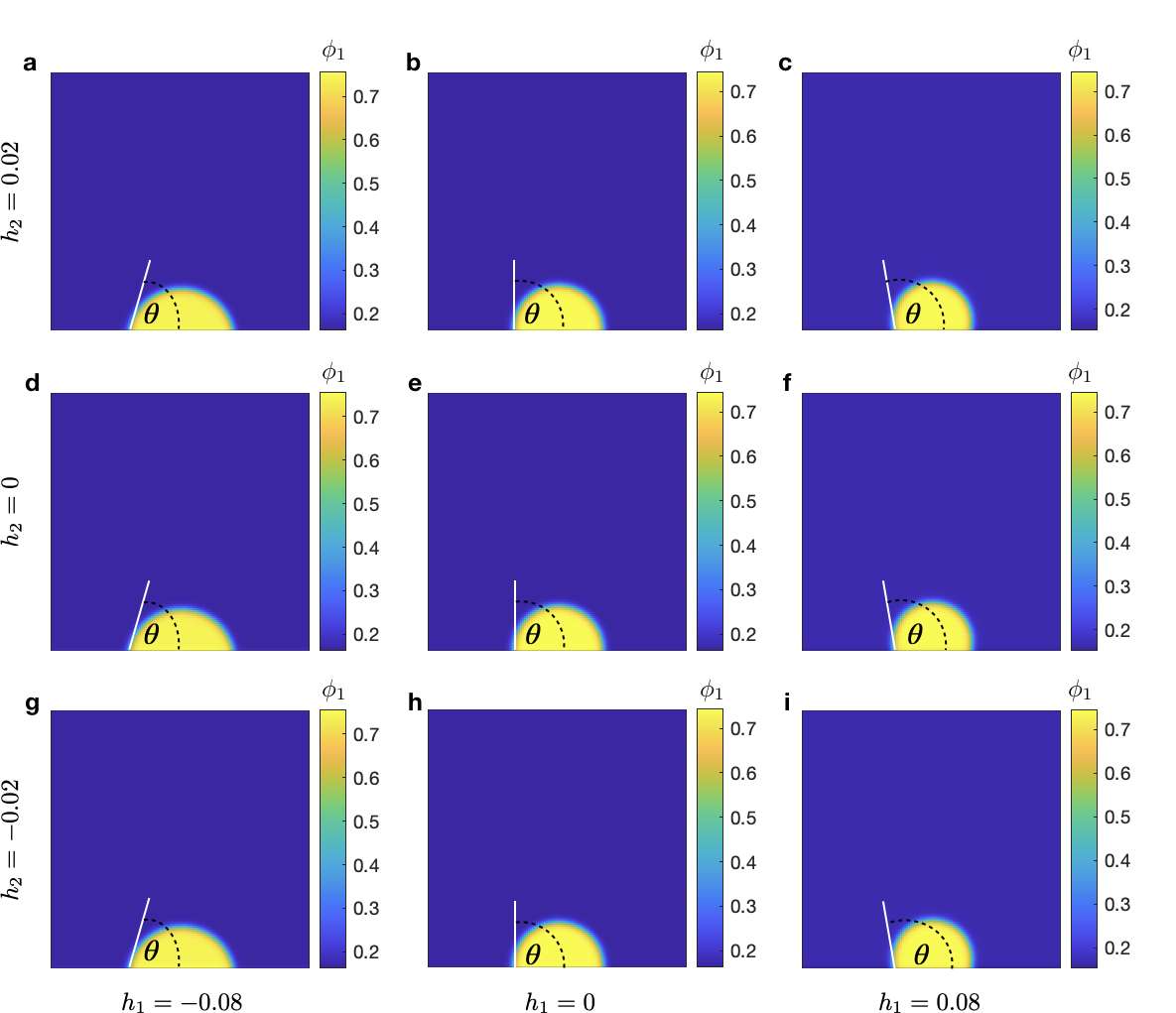}}
\caption{\textbf{Profiles of $\phi_1$ with {a} linear interaction between solid walls and mixture components on the contact angle of a {two-phase} coexisting mixture {in equilibrium}}. The snapshots {denote} the profiles of component 1, i.e. $\phi_1$ with different values of $h_1$ and $h_2$. At each snapshot, the yellow area depicts the condensates, a white line is used to denote the tangent line of the spherical interface at contact point on the surface $\Gamma$. The contact angle of droplet between tangent line and surface is {denoted} by $\theta$.  In each column (row), we use same $h_1$($h_2$) values. The basic parameter values used in this case {are}: $\kappa_{1\Gamma} = \kappa_{2\Gamma}=0$, $\kappa_1= \kappa_2 = 1$, $g_1 = g_2 = \gamma = 0$, $\chi_{12} = -1, \chi_{23} = 0, \chi_{13} = 2.5$. Since {the} droplet is composite of $\phi_1$ mainly, the contact angle is determined by the value of $h_1$ in this study. }
\label{fig:h12_phi_A}
\end{figure}

\begin{figure}
\centering
{\includegraphics[width=1\textwidth]{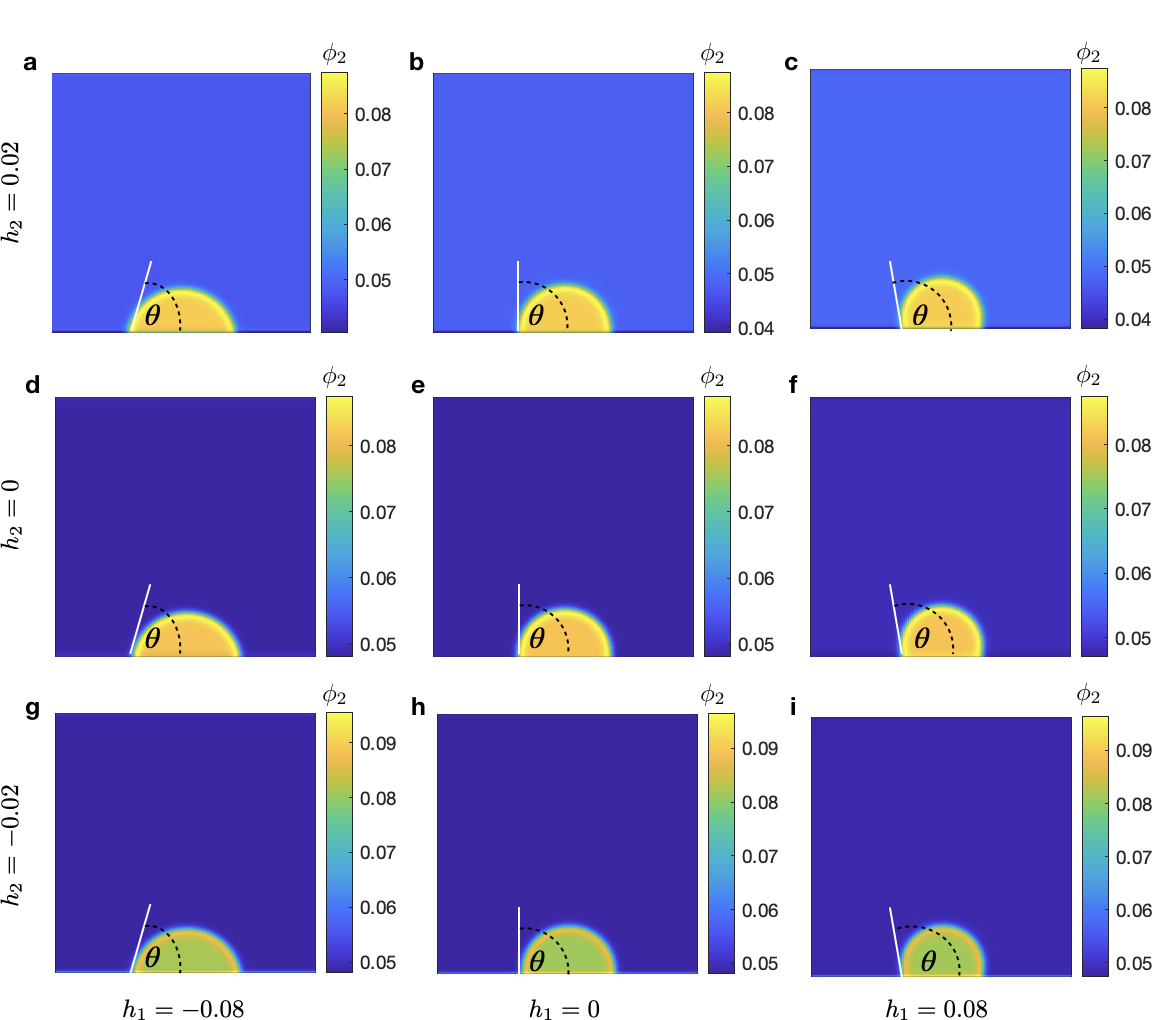}}
\caption{\textbf{Profiles of $\phi_2$ with {a} linear interaction between solid walls and mixture components on the contact angle of a {two-phase} coexisting mixture {in equilibrium}}. The snapshots {denote} the profiles of component 2, i.e. $\phi_2$ with different values of $h_1$ and $h_2$. The profile of $\phi_2$ changes rapidly near by the surface with respect to different $h_2$ values. However, the contact angle is determined by the value of $h_1$ since the the droplet mainly contains $\phi_1$. The basic parameter values used in this case {are}: $\kappa_{1\Gamma} = \kappa_{2\Gamma}=0$, $\kappa_1= \kappa_2 = 1$, $g_1 = g_2 = \gamma = 0$, $\chi_{12} = -1, \chi_{23} = 0, \chi_{13} = 2.5$. }
\label{fig:h12_phi_R}
\end{figure}

% Effects of $g_i$ on the wettability stationary solution.
% Figure 5
% \com{results check:  nine h5 files}

\begin{figure}
\centering
{\includegraphics[width=1\textwidth]{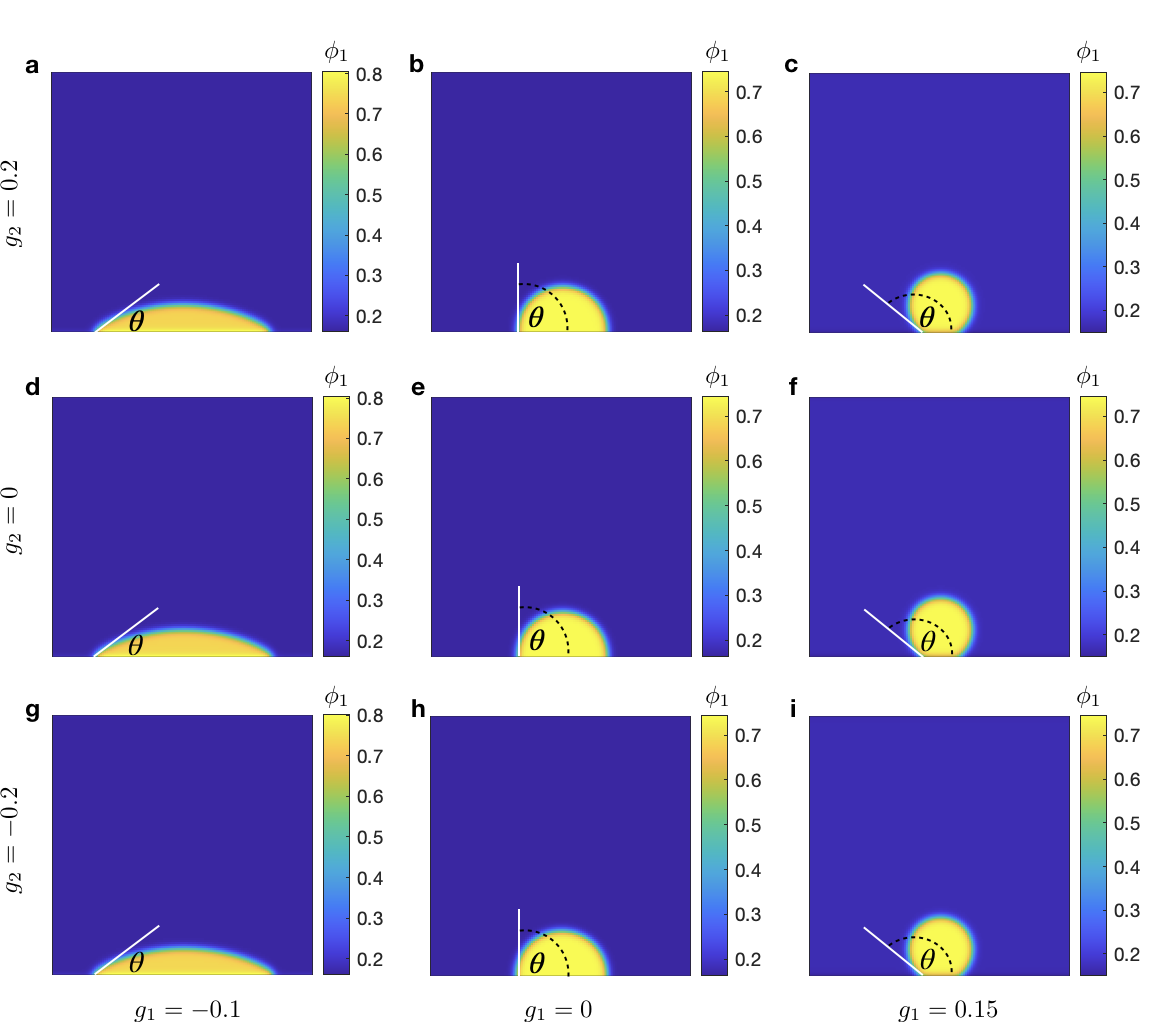}}
\caption{\textbf{Profiles of $\phi_1$ with {a} quadratic interaction between solid walls and mixture components on the contact angle of a {two-phase} coexisting mixture {in equilibrium}}. The snapshots {denote} the profiles of component 1, i.e. $\phi_1$ with different values of $g_1$ and $g_2$. We observe that the interaction between $\phi_2$ and solid wall influences the contact angle mainly. The basic parameter values used in this case {are}: $\kappa_{1\Gamma} = \kappa_{2\Gamma}=0$, $\kappa_1= \kappa_2 = 0$, $h_1 = h_2 = \gamma = 0$, $\chi_{12} = -1, \chi_{23} = 0, \chi_{13} = 2.5$. }
\label{fig:g12_phi_A}
\end{figure}

\begin{figure}
\centering
{\includegraphics[width=1\textwidth]{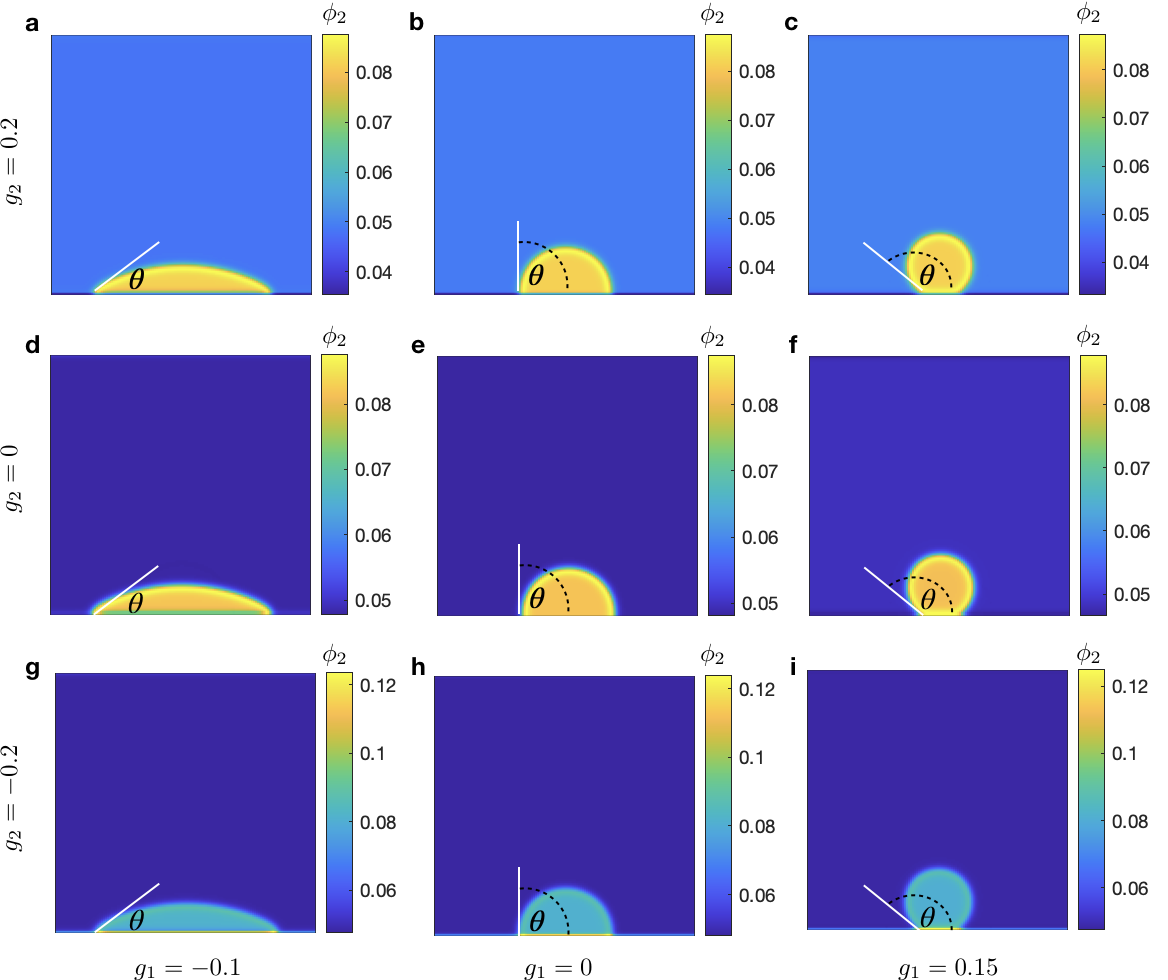}}
\caption{\textbf{Profiles of $\phi_2$ with {a} quadratic interaction between solid walls and mixture components on the contact angle of a {two-phase} coexisting mixture {in equilibrium}}. The snapshots {denote} the profiles of component 2, i.e. $\phi_2$ with different values of $g_1$ and $g_2$. We find that the interaction between $\phi_1$ and solid surface changes the profiles of $\phi_2$ near by the surface, while the contact angle is determined by interaction between $\phi_1$ and solid wall. The basic parameter values used in this case {are}: $\kappa_{1\Gamma} = \kappa_{2\Gamma}=0$, $\kappa_1= \kappa_2 = 1$, $h_1 = h_2 = \gamma = 0$, $\chi_{12} = -1, \chi_{23} = 0, \chi_{13} = 2.5$. }
\label{fig:g12_phi_R}
\end{figure}

\begin{figure}
\centering
{\includegraphics[width=1\textwidth]{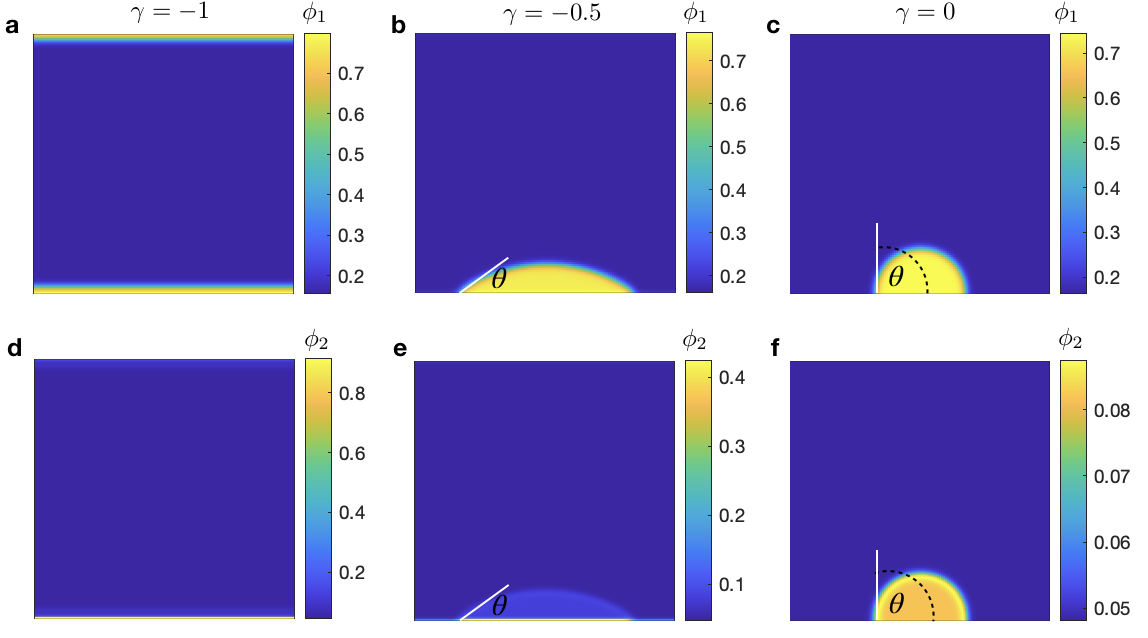}}
\caption{\textbf{Profiles of $\phi_1$ and $\phi_2$ with {a} nonlinear coupling interaction between $\phi_1$ and $\phi_2$ {in} equilibrium}. {Plots (\textbf{a}, \textbf{b}, \textbf{c})} represent the concentration profiles of $\phi_1$, while {plots (\textbf{d}, \textbf{e}, \textbf{f})} the concentration profiles of $\phi_2$. We observe that, as the negative coupling interaction between $\phi_1$ and $\phi_2$ becomes {stronger}, the contact angle decreases until the condensate (yellow) completely wets on the solid surface (plots {(\textbf{a}, \textbf{d})}. Here, we only investigate the negative coupling {interaction}, since the positive coupling {interaction} will lead to nonphysical solutions. The basic parameter values used in this case are: $\kappa_{1\Gamma} = \kappa_{2\Gamma} = 0$, $\kappa_1= \kappa_2 = 1$, $h_1 = h_2 = g_1 = g_2 = 0$, $\chi_{12} = -1, \chi_{23} = 0, \chi_{13} = 2.5$. } 
\label{fig:gamma_phi_a}
\end{figure}

\begin{figure}
\centering
{\includegraphics[width=1\textwidth]{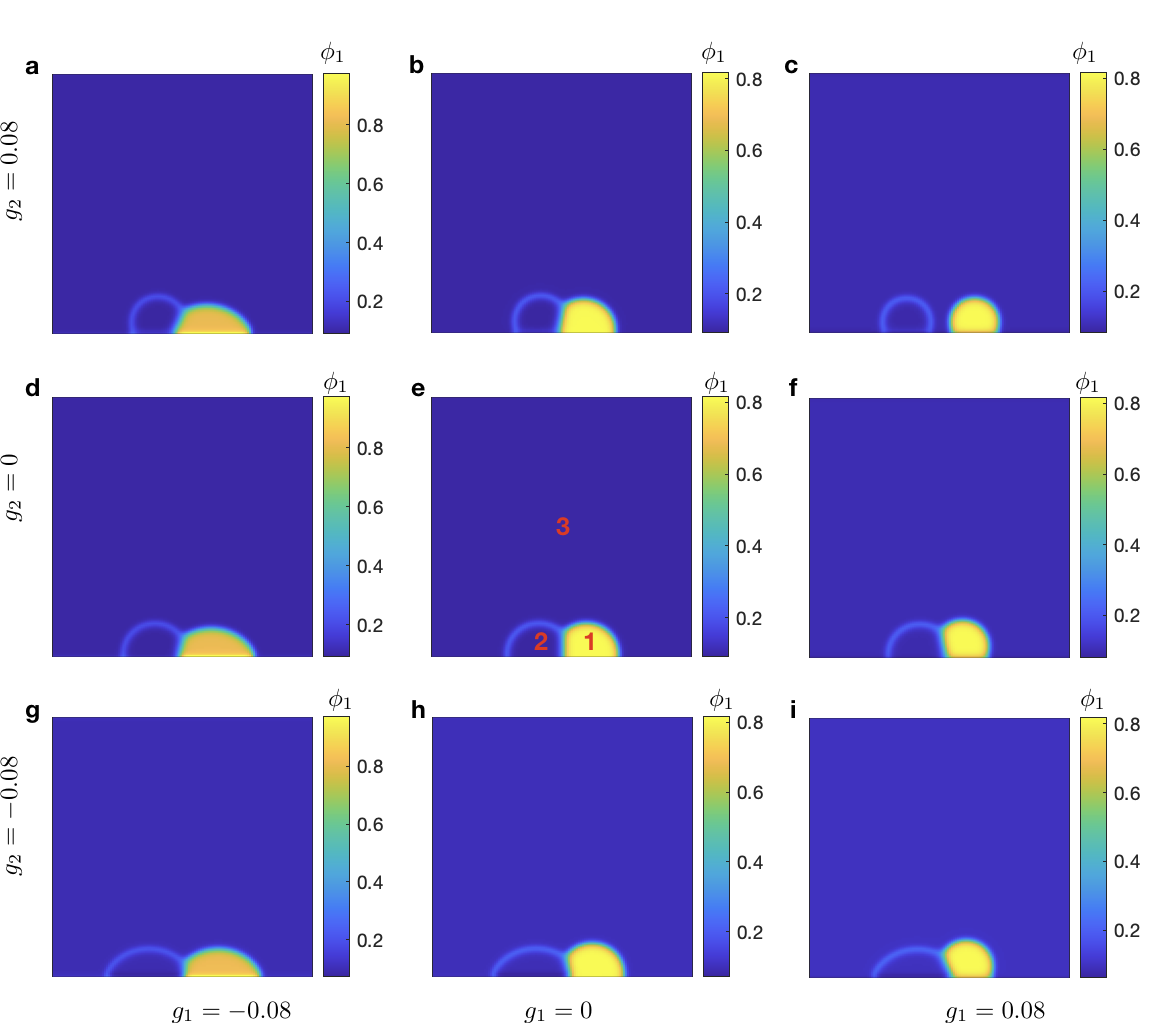}}
\caption{\textbf{Profiles of $\phi_1$ with {a} quadratic interaction between solid wall and mixture components on the contact angle of a {three-phase} coexisting mixture at equilibrium}. With the interaction parameter values $\chi_{12} = \chi_{23} = \chi_{13} = 3$, the ternary mixture has three coexisting phases (denoted by red numbers 1, 2, 3 in plot {\textbf{e}}). We change the strength of {the} quadratic interaction, i.e. the values of $g_i, i = 1,2$. We observe that the contact angle of droplet 1 and 2 are mainly determined by the interaction between the dominant component and the solid wall, i.e. $\phi_1$ in droplet 1 and $\phi_2$ in droplet 2. Other basic parameter values used in this case are: $\kappa_{1\Gamma} = \kappa_{2\Gamma} = 0$, $\kappa_1= \kappa_2 = 1$, $h_1 = h_2 = \gamma = 0$.} 
\label{fig:three_phases_g12_phi_A}
\end{figure}

\begin{figure}
\centering
{\includegraphics[width=1\textwidth]{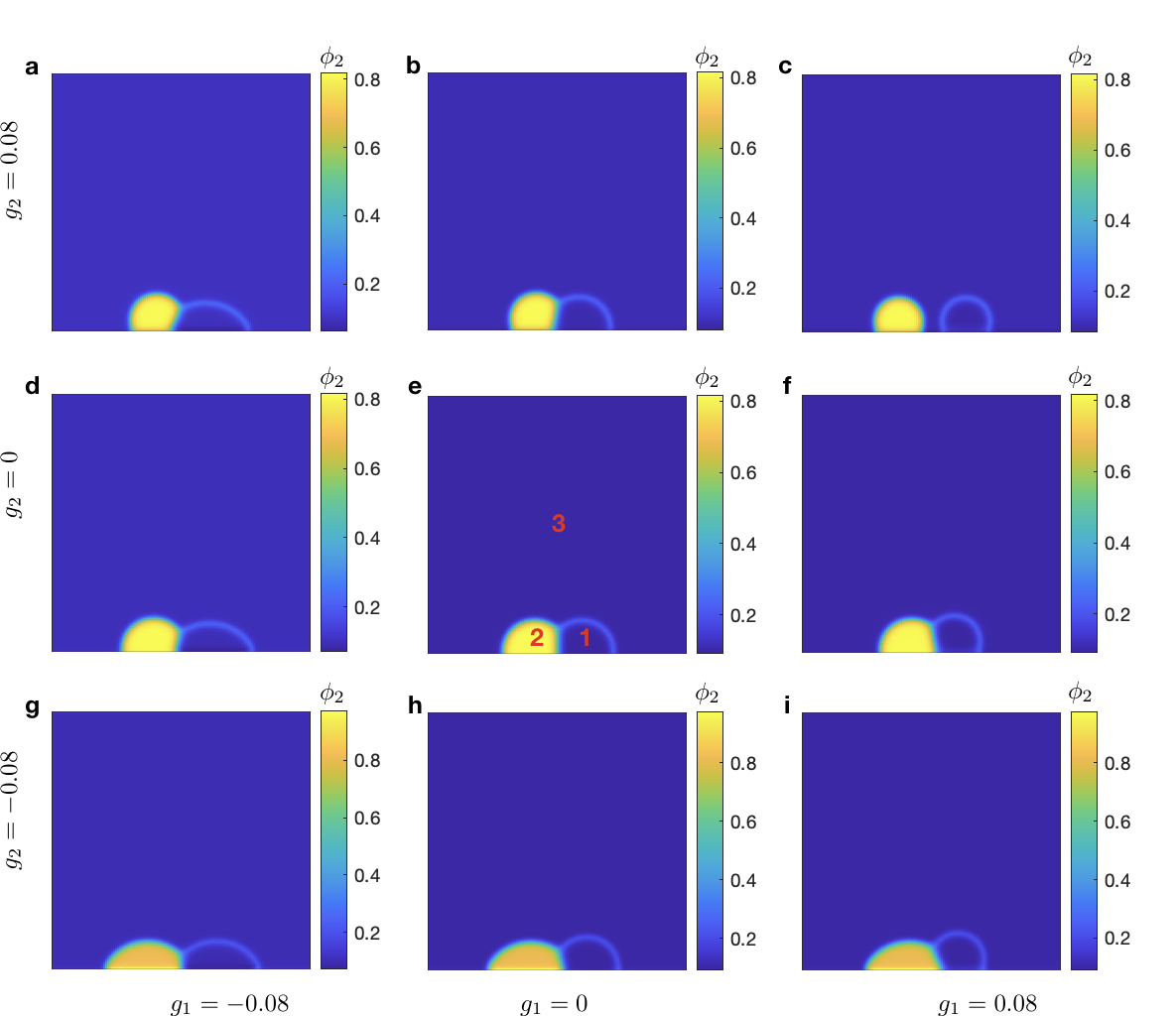}}
\caption{\textbf{Profiles of $\phi_1$ with {a} quadratic interaction between solid wall and mixture components on the contact angle of a {three-phase} coexisting mixture at equilibrium}. The basic parameter values used in this case {are}: $\kappa_{1\Gamma} = \kappa_{2\Gamma} = 0$, $\kappa_1= \kappa_2 = 1$, $h_1 = h_2 = \gamma = 0$, $\chi_{12} = \chi_{23} = \chi_{13} = 3$.} 
\label{fig:three_phases_g12_phi_R}
\end{figure}

\begin{figure}
\centering
{\includegraphics[width=1\textwidth]{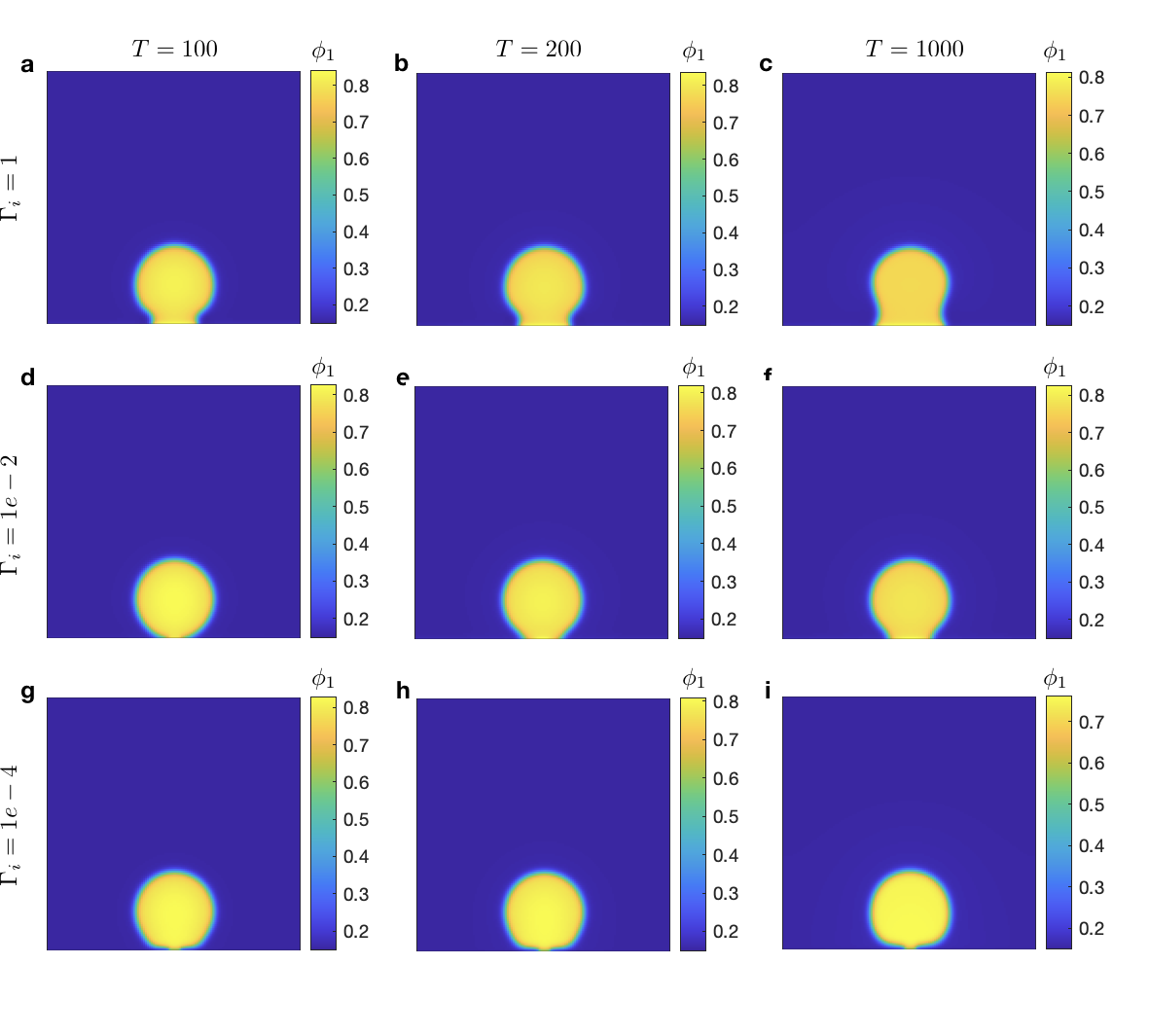}}
\caption{\textbf{Relaxation of dynamic boundary conditions controls the droplet spreading on the surface}. Plots {(\textbf{a}-\textbf{i})} are the snapshots of component 1, i.e. $\phi_1$. We do not consider the cross-coupling interaction in this study, i.e. $\Gamma_{12} = 0$. The relaxation rate $\Gamma_i, i=1, 2$ decrease from 1 to $10^{-4}$, the spreading process slows down correspondingly. Other basic parameter values used in this case are: {$\kappa_{1\Gamma} = \kappa_{2\Gamma}=0$, $\kappa_1= \kappa_2 = 1$, $h_1 = 0, h_2 = 0, g_1 = -1,  g_2 =0, \gamma = 0$, $\chi_{12} = -1$, $\chi_{23} = 0$, $\chi_{13} = 2.5$.} }
% \SL{do we need the values of $h_1$ and $h_2$?} 
\label{fig:spreading_panel1}
\end{figure}

\begin{figure}
\centering
{\includegraphics[width=1\textwidth]{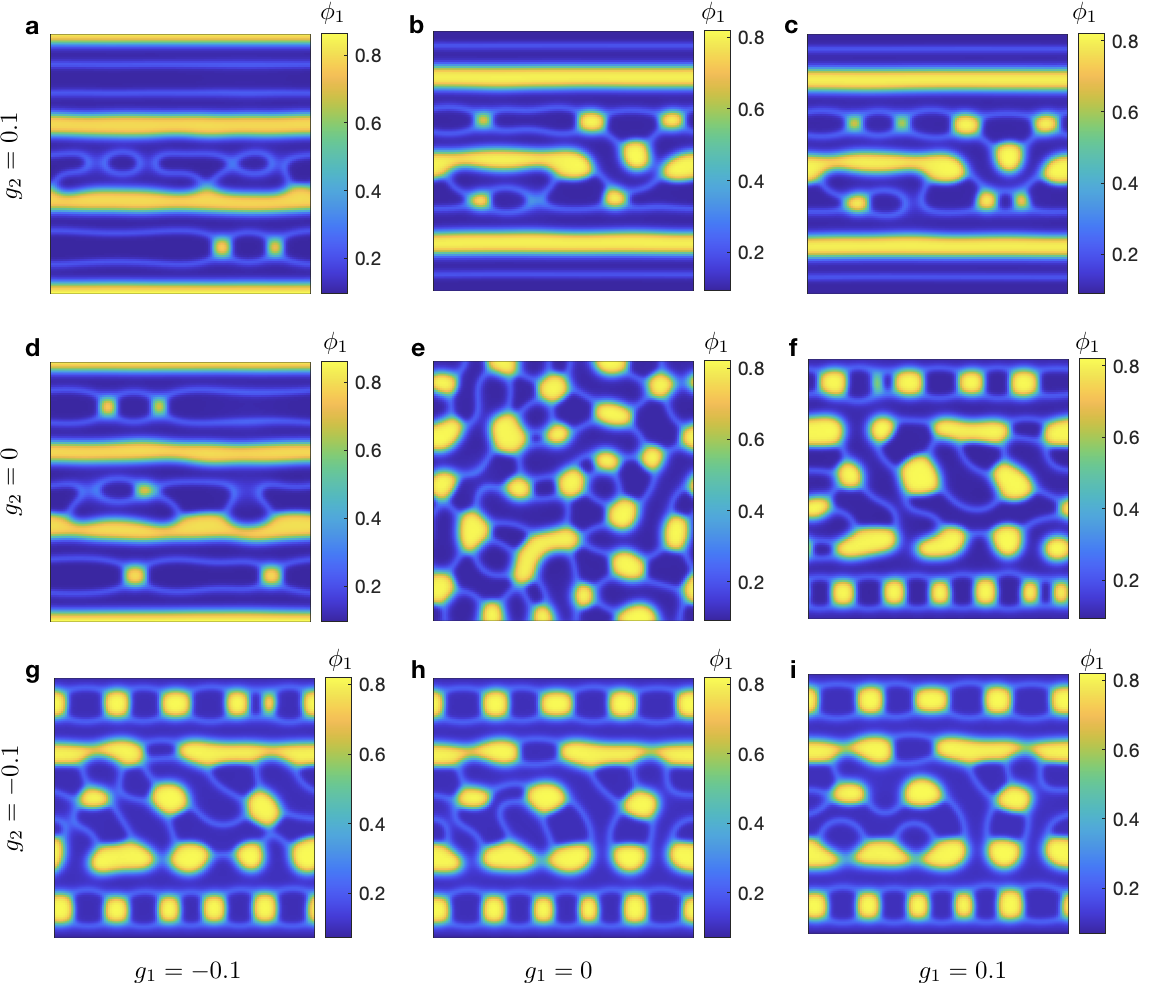}}
\caption{\textbf{Profiles of $\phi_1$ during the spontaneous phase separation of ternary mixture with three-phase coexistence with various {types} of wall-mixture interactions}. We show the snapshots of $\phi_1$ at $T=10000$. With neutral wall-mixture interaction (plot {\textbf{e}}, i.e. $g_i = 0, i=1,2.$), the condensates (yellow) distribute in the bulk homogeneously. However, with attractive/repulsive wall-mixture interaction, the condensates distribute horizontally parallel with the solid wall {(see plot \textbf{a},\textbf{b}, \textbf{c}, \textbf{d}, \textbf{f}, \textbf{g}, \textbf{i})}. The basic parameter values used in this case are: {$h_1 = h_2 = 0$}, $\kappa_{1\Gamma} = \kappa_{2\Gamma}=1$, $\kappa_1= \kappa_2 = 1$, $\gamma = 0$, $\chi_{12} = \chi_{23} = \chi_{13} = 3$. } 
\label{fig:PS_g12_phi_A}
\end{figure}

\begin{figure}
\centering
{\includegraphics[width=1\textwidth]{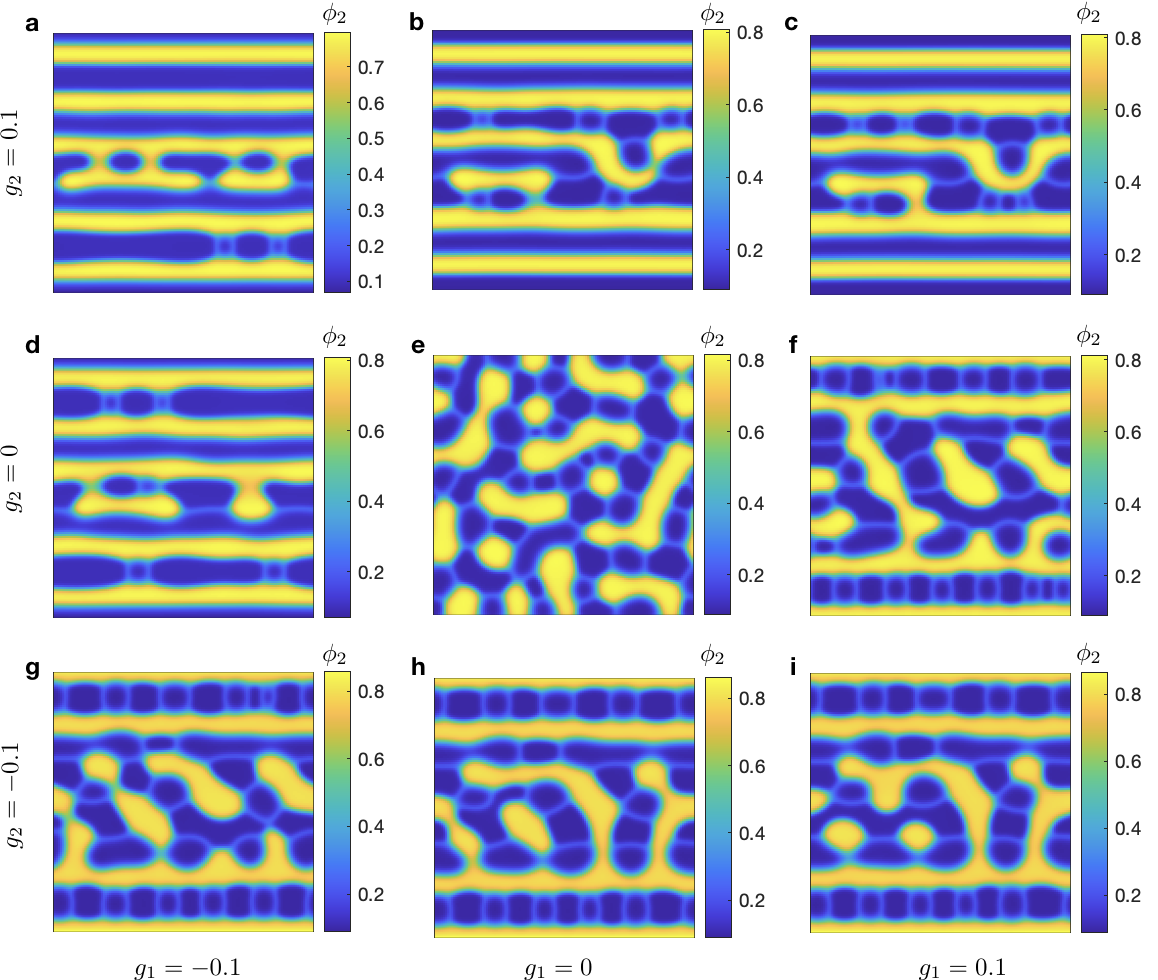}}
\caption{\textbf{Profiles of $\phi_2$ during the spontaneous phase separation of ternary mixture with three-phase coexistence with various {types} of wall-mixture interactions}. The basic parameter values used in this case {are}: {$h_1 = h_2 = 0$}, $\kappa_{1\Gamma} = \kappa_{2\Gamma}=1$, $\kappa_1= \kappa_2 = 1$, $ \gamma = 0$, $\chi_{12} = \chi_{23} = \chi_{13} = 3$.  } 
\label{fig:PS_g12_phi_R}
\end{figure}

\begin{figure}
\centering
{\includegraphics[width=1\textwidth]{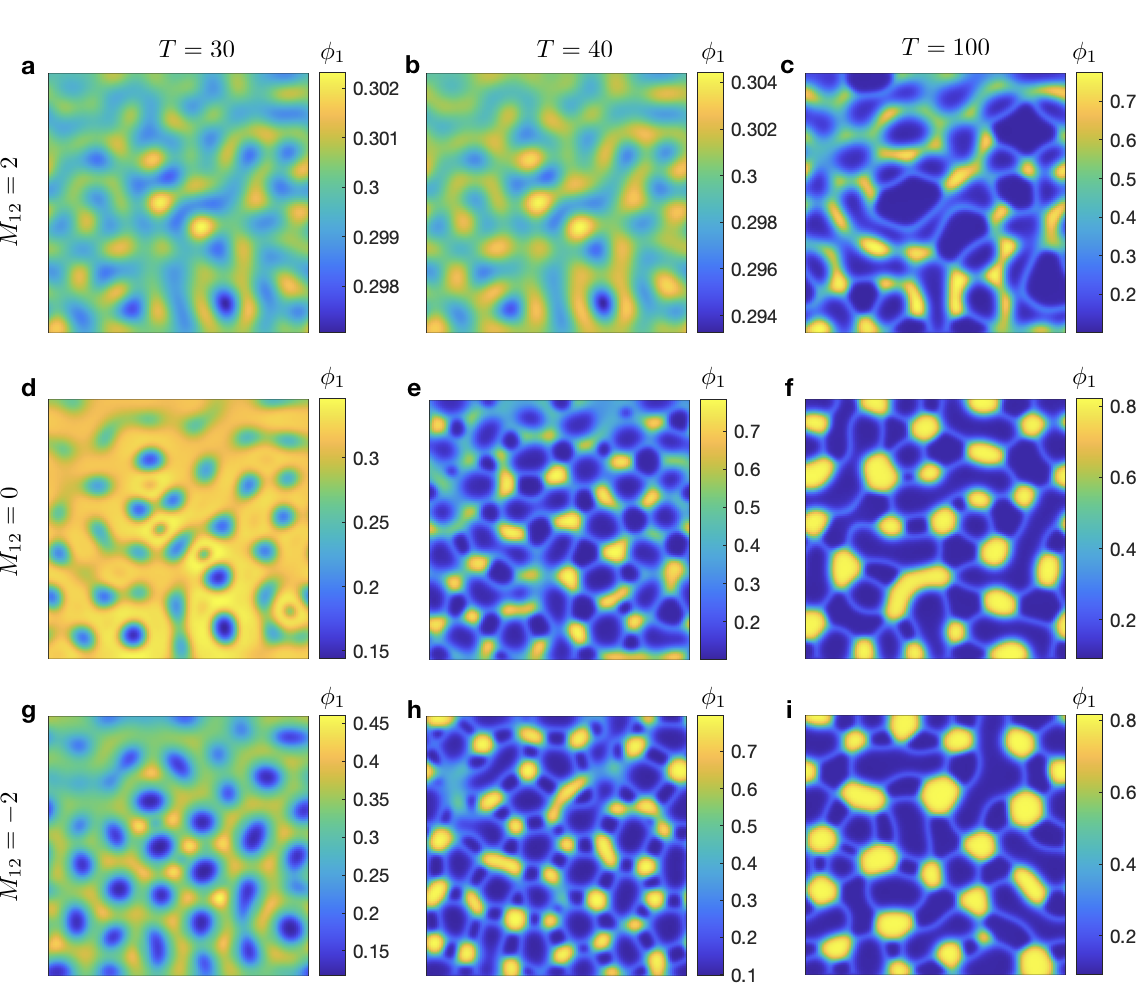}}
\caption{\textbf{Profiles of $\phi_1$ during spontaneous phase separation evolution at different time points with different {values of} cross coupling coefficient $M_{12}$}. We find that changing the cross-coupling coefficient {$M_{12}$} {leads} to changes in the phase separation and following coarsening processes. The basic parameter values used in this study are: {$h_1 = h_1 = g_1 = g_2 = \gamma = 0$}, $\kappa_{1\Gamma} = \kappa_{2\Gamma}=1$, $\kappa_1= \kappa_2 = 1$, $ \gamma = 0$, $\chi_{12} = \chi_{23} = \chi_{13} = 3$.} 
\label{fig:M12_phi_A}
\end{figure}

\begin{figure}
\centering
{\includegraphics[width=1\textwidth]{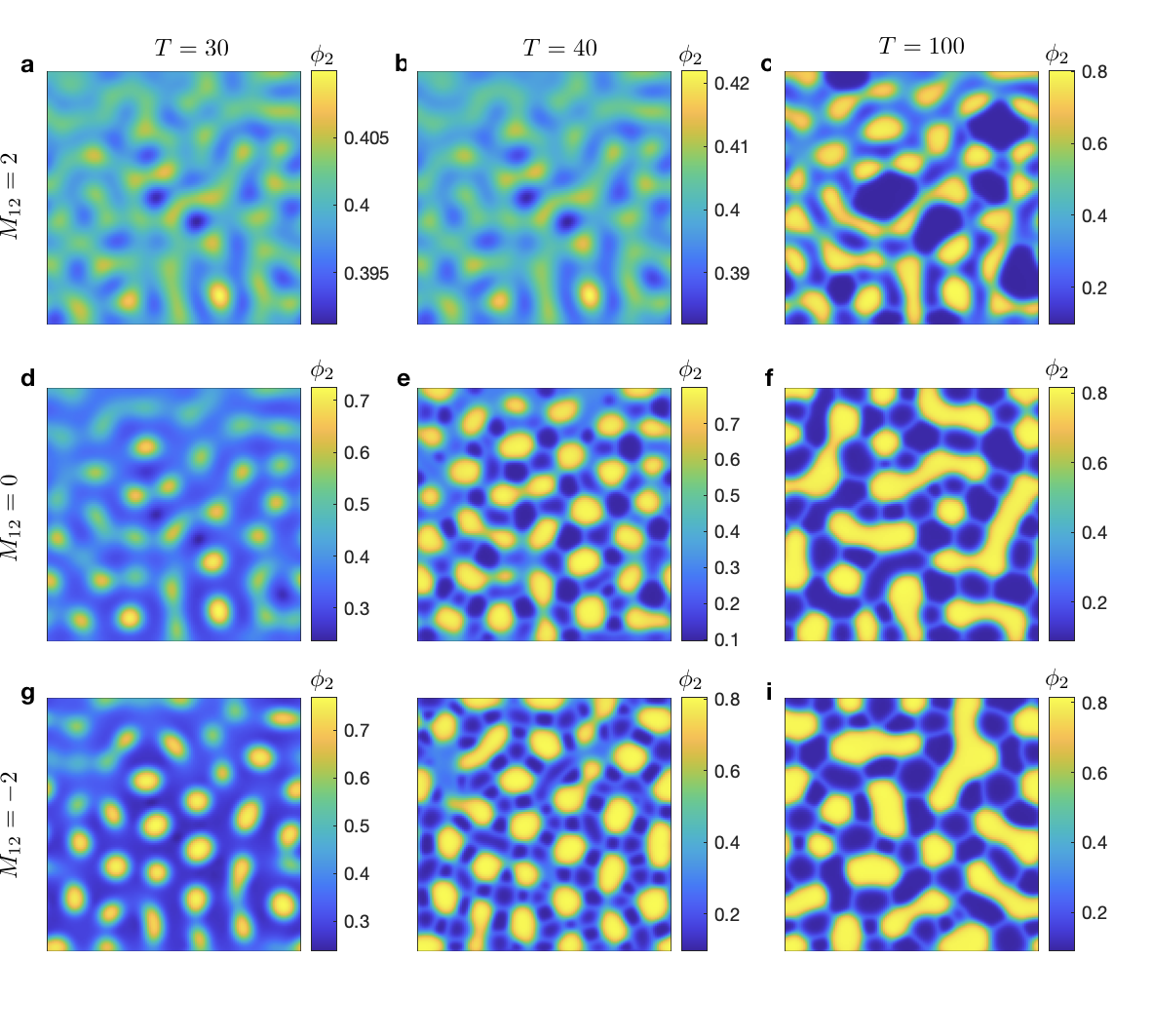}}
\caption{\textbf{Profiles of {$\phi_2$} during spontaneous phase separation evolution at different time points with different {values of} {cross-coupling} coefficient $M_{12}$}. The basic parameter values used in this case {are}: {$h_1 = h_1 = g_1 = g_2 = \gamma = 0$}, $\kappa_{1\Gamma} = \kappa_{2\Gamma}=1$, $\kappa_1= \kappa_2 = 1$, $ \gamma = 0$, $\chi_{12} = \chi_{23} = \chi_{13} = 3$.} 
\label{fig:M12_phi_R}
\end{figure}

\begin{figure}
\centering
{\includegraphics[width=1\textwidth]{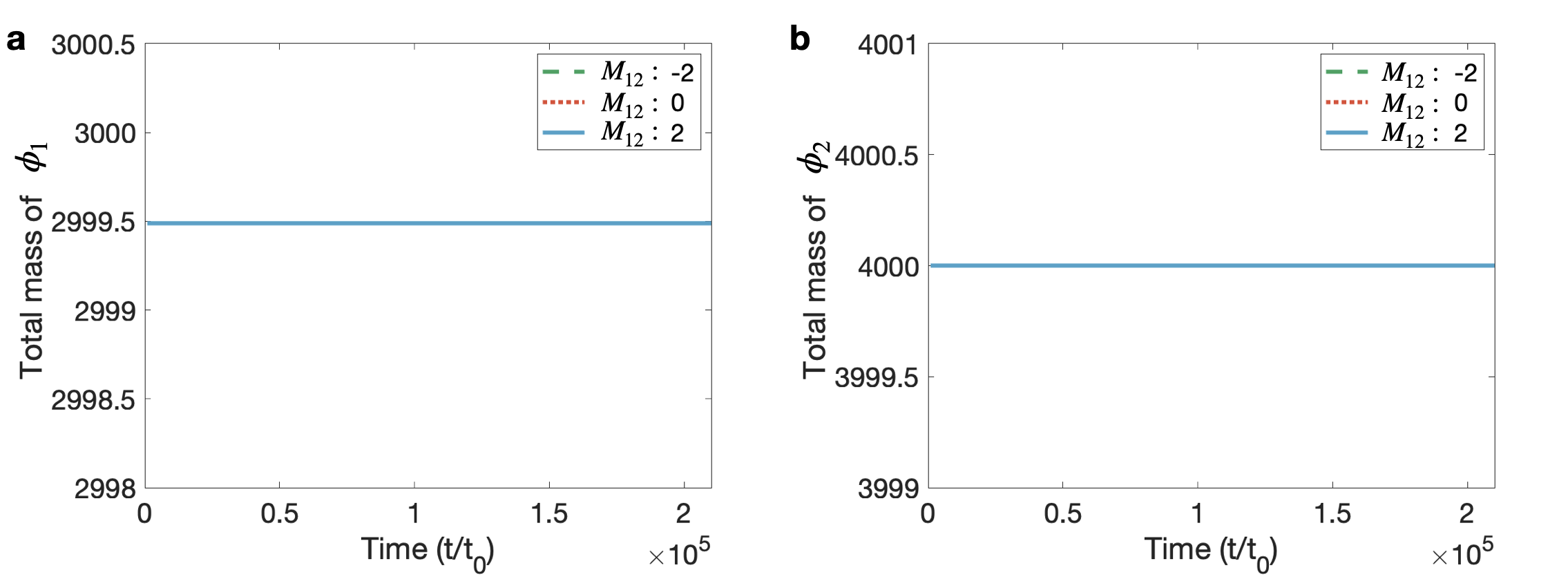}}
\caption{\textbf{Total mass of $\phi_1$ and $\phi_2$ in the system during spontaneous phase separation process.} The basic parameter values used in this case {are}: {$h_1 = h_1 = g_1 = g_2 = \gamma = 0$}, $\kappa_{1\Gamma} = \kappa_{2\Gamma}=1$, $\kappa_1= \kappa_2 = 1$, $ \gamma = 0$, $\chi_{12} = \chi_{23} = \chi_{13} = 3$.} 
\label{fig:M12_Mass}
\end{figure}

\begin{figure}
\centering
{\includegraphics[width=0.6\textwidth]{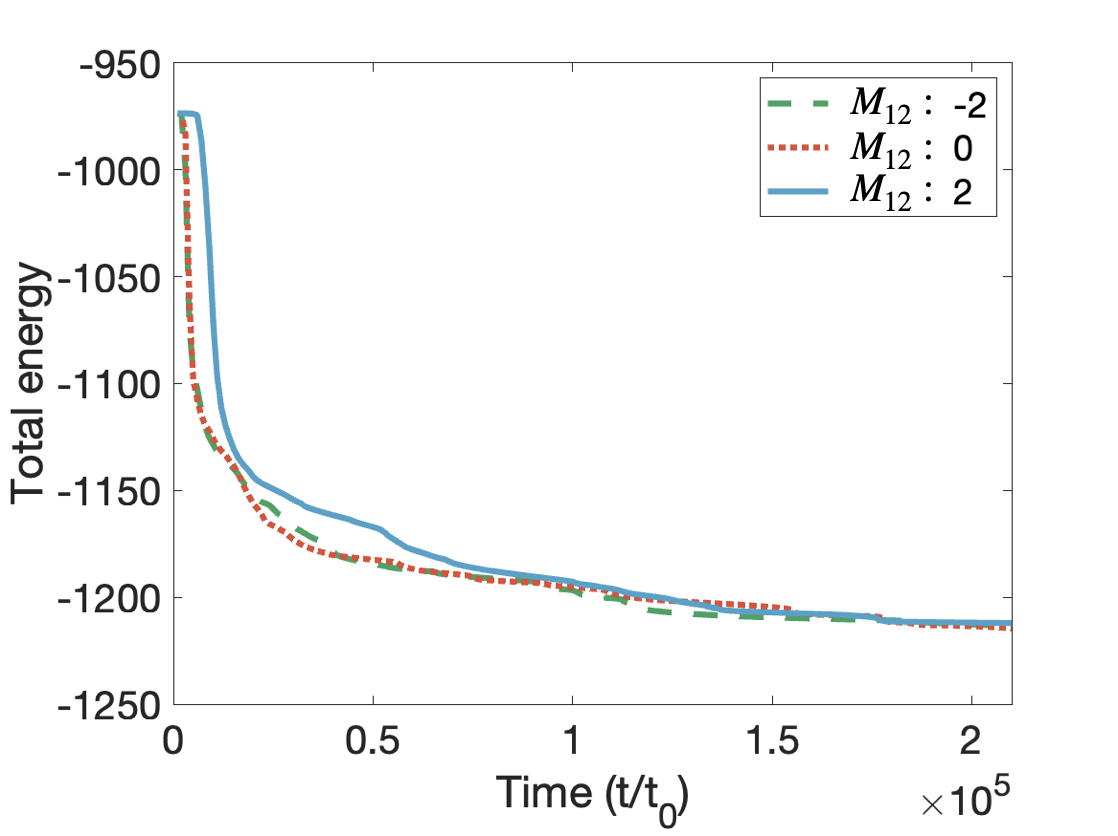}}
\caption{\textbf{Total energy profiles  during spontaneous phase separation process with different cross-coupling coefficient $M_{12}$.} The basic parameter values used in this case {are}: {$h_1 = h_1 = g_1 = g_2 = \gamma = 0$}, $\kappa_{1\Gamma} = \kappa_{2\Gamma}=1$, $\kappa_1= \kappa_2 = 1$, $ \gamma = 0$, $\chi_{12} = \chi_{23} = \chi_{13} = 3$.} 
\label{fig:M12_Energy}
\end{figure}

\section{Conclusions}\label{sec:conclusions}

In this paper, we have presented a systematic derivation of multi-component multi-phase mixtures with dynamic boundary conditions. The governing equations in the models are composed of the mass conservation as well as the constitutive equations, which are derived using the Onsager principle to preserve the energy dissipation rate in time. The gradient flow structure is $H^{-1}(\Omega)$ in the bulk and $L_2(\Gamma)$ on the surface.

Moreover, we have presented a second order, fully--discrete, linear and unconditionally energy stable numerical scheme for the ternary mixture model with dynamic boundary conditions. Firstly, we reformulate the model by introducing an intermediate variables following the Energy Quadratization strategy. Using the reformulated model equations, we develop a second order, energy stable, semi-discrete numerical scheme in time. Then, we obtain a fully--discrete numerical scheme by applying the finite difference method on the staggered grid in space, which preserves a fully discrete energy dissipation rate and total mass conservation laws.

% \com{rewrite this paragraph}
{Beyond theoretical advancements, our work has substantial engineering implications.} Utilizing our efficient and accurate numerical solver, we investigate the effects of short-ranged interaction between mixture components and solid surface on the wettability of the surface, from the perspectives of both stationary and kinetic solutions. We find that the contact angle of condensates are determined by the additive effects introduced by the interaction between each component and solid wall. The droplet spreading can be affected hugely by the surface relaxation rates, while the cross-coupling relaxation rate in the surface does not influence the droplet spreading a lot. 
Furthermore, the spontaneous phase separation phenomena of ternary mixture with two phases and three phases, with dynamic boundary conditions are explored. Our results suggest that strong attractive/repulsive coupling between surface and bulk {leads} to ordered condensates patterns in the bulk. {Moreover}, the cross-coupling in the bulk {changes} the phase separation process significantly. {This insight is invaluable in fields like semiconductor manufacturing, where surface properties are critical, or in medical device fabrication, where understanding biofluid interactions with surfaces can lead to better product designs.}
% {These findings could revolutionize approaches in inkjet printing technologies or improve methods in pesticide delivery in agriculture.}
% \SL{The effects of wall-mixture interaction and dynamic boundary conditions to the multi-component multi-phase mixture are tested based on our ternary mixture model and efficient numerical scheme, specially in wettability, spontaneous phase separation, and coarsening processes. Key findings are summarized as the following: 
% (1) The contact angle of condensates are determined by the additive effects introduced by the interaction between each component and solid wall. The droplet spreading can be affected hugely by the surface relaxation rates, while the cross-coupling relaxation rate in the surface does not influence the droplet spreading a lot.
% (2) The strong attractive/repulsive coupling between surface and bulk leads to ordered condensates patterns in the bulk. Furthermore, the cross-coupling in the bulk changes the phase separation process significantly.}

{Furthremore, our} model and the scheme can be readily extended to models of N-component mixtures with $N > 3$. Our work provides a general framework for the study of multi-component mixture with dynamic boundary conditions. The multi-component mixture with more complex boundary conditions(DBC), e.g. Goldstein DBC \eqref{eq:DBC2}, Liu-Wu model DBC \eqref{eq:DBC3} and KLLM model DBC \eqref{eq:DBC4} can be studied based on this framework. 
{This work not only provides a foundational framework for future study of multi-component mixtures with dynamic boundary conditions for a wide range of applications but also makes a significant influence in the study of complex fluid dynamics.}

 \section*{Acknowledgement}
 {
We thank C. A. Weber and Z.\ Zhang for insightful discussions. We also thank the anonymous referees for their critical comments. 
}

\section{Appendix}
\subsection{Phase diagram of ternary mixture}\label{sec:appendix1}

{
We use convex hull algorithm to calculate the phase diagram of the ternary mixture. The interaction parameters values are $\chi_{12} = \chi_{23} = \chi_{13} = 3$, the wall-mixture interaction coefficients: $h_1 = h_2 = g_1 = g_2 = \gamma = 0$. }

\begin{figure}
\centering
{\includegraphics[width=1\textwidth]{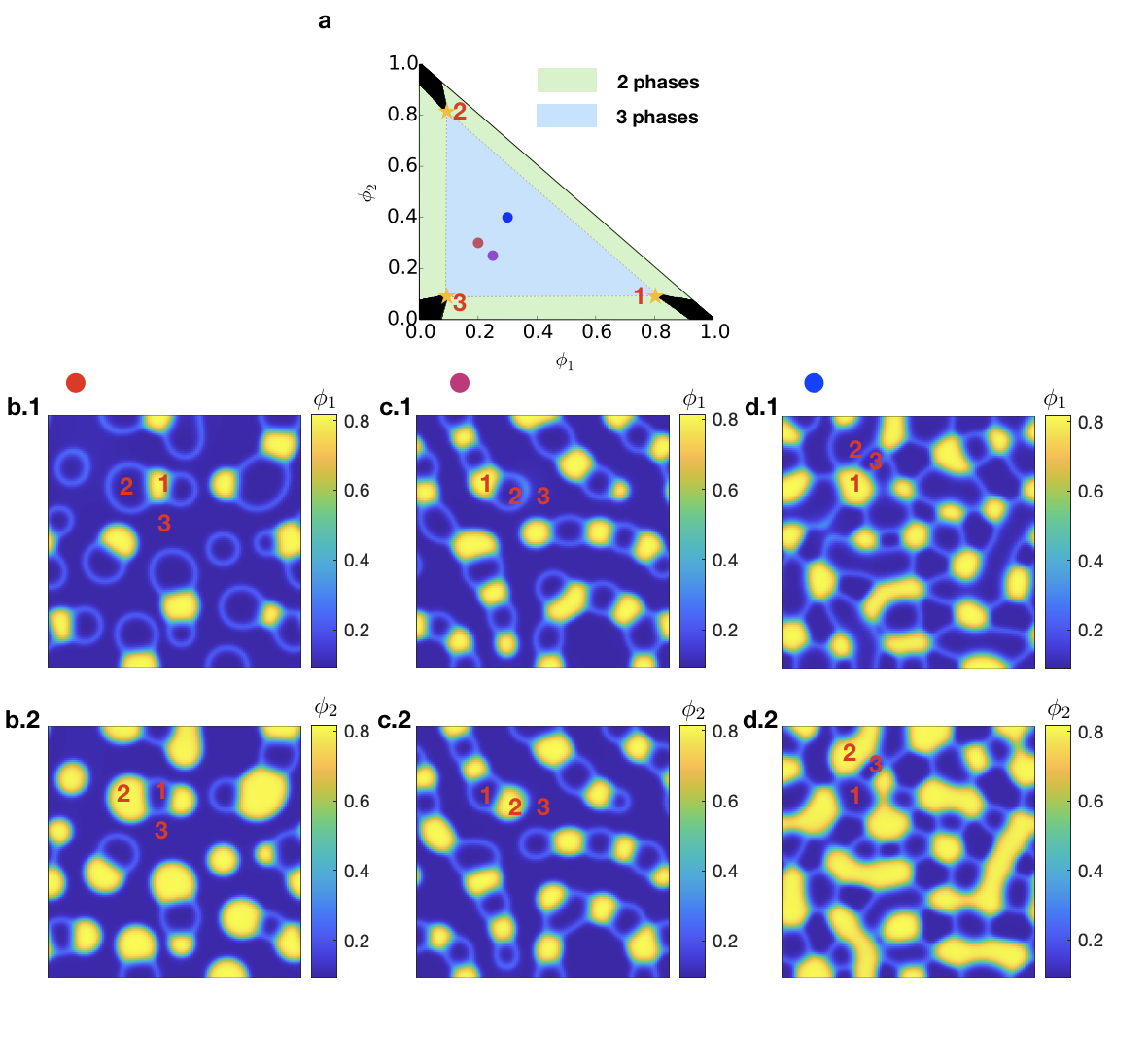}}
\caption{\textbf{Phase diagram of ternary mixture and three-phase coexistence examples}. Plot {\textbf{a}} is the phase diagram of ternary mixture. Black region depicts the case with one homogeneous phase; Green region depicts the case with two phases; Blue region denotes the case with three coexisting phases. Plots {\textbf{b.1}-\textbf{b.2}, \textbf{c.1}-\textbf{c.2}, \textbf{d.1}-\textbf{d.2}} are snapshots of $\phi_1$ and $\phi_2$ with different {sets} of {the initial composition}, which are marked by the color dots in the blue region. The basic parameter values used in this case are: {$h_1 = 0, h_2 = 0$,} $\kappa_1 = \kappa_2 = 1$, $g_1 = g_2 = \gamma = 0$, $\chi_{12} = \chi_{23} = \chi_{13} = 3$. } 
\label{fig:ternary_phase_diagram}
\end{figure}

In the phase diagram Fig~\ref{fig:ternary_phase_diagram}, the x-axis represents the composition of one of the components $\phi_1$ in the mixture, while the y-axis represents the composition of another component $\phi_2$. The concentration of third component equals to $1-\phi_1-\phi_2$ based on the incomprehensibility assumption. This phase diagram depicts the behavior of ternary mixture at equilibrium. 
The black regions represent the mixture which has no phase separation at equilibrium. The ternary mixture which has the compositions at light green regions will phase separate into two coexisting phases with different composition profiles. In the blue region, there are three coexisting phases. The composition of these three phases are denoted by the yellow stars in the three corners of the region.
Here, we give three examples of three phase coexisting ternary mixture which are depicted by three colorful dots(red, magenta, blue). Correspondingly, in the Fig.~\ref{fig:ternary_phase_diagram}-(b.1-b.2, c.1-c.2, d.1-d.2), we show the three phases coexisting system with phases 1, 2, 3. In specific, the Fig.~\ref{fig:ternary_phase_diagram}-(b.1, c.1, d.1) denotes the concentration profiles of $\phi_1$ in each case, while Fig.~\ref{fig:ternary_phase_diagram}-(b.2, c.2, d.2) the concentration profiles of $\phi_2$.
All the numerical studies of three-phases coexisting phenomena in the main text are based on this phase diagram.

% \newpage

\subsection{Abbreviations and Notations}
\begin{table}[h]
\captionsetup{justification=centering}
\caption{Abbreviations}
\renewcommand\arraystretch{1.5}
\centering
\begin{tabular}{c c}
\hline
\textbf{Abbreviations} & \textbf{Explanations} \\
\hline
DBC & Dynamic Boundary Condition \\
IEQ & Invariant Energy Quadratization \\
\hline
\end{tabular}
\end{table}

%%%%%%%%%%%%%%%%%%%%$
\begin{table}[h]
\captionsetup{justification=centering}
\caption{Notations}
\renewcommand\arraystretch{1.5}
\centering
\begin{tabular}{c c}
\hline
\textbf{Notations} & \textbf{Explanations} \\
\hline
$\chi_{ij}$ & interaction strength between the component i and j \\
$\gamma$ & coupling interaction strength between components on the surface \\
$\Gamma$ & boundary of domain \\
$\Gamma_i$ & relaxation parameters of dynamic boundary condition, i=1,2,12 \\
% $\Gamma_s$ & surface kinetic coefficient \\
$\kappa_B$ & Boltzmann constant \\
$\kappa_i$,$\kappa_{i\Gamma}$ & gradient coefficients associated with interface free energies, i=1,2,12 \\
$\mu_i$ & chemical potentials in the bulk for components i=1,2 \\
$\mu_{i\Gamma}$ & chemical potentials on the surface for components i=1,2 \\
$\Delta_\Gamma$ & Laplace–Beltrami operator on the boundary surface  \\
% $\nu$ & molecule volume of component 3 \\
$\nu_i$ & molecule volume of component i, i=1,2 \\
$\Omega$ & domain \\
$\omega_i$ & internal free energy coefficient for component i in the bulk \\
$\phi_i$ & volume fractions of three components, i=1,2,3 \\
$\phi_{i\Gamma}$ & volume fractions of component, i=1,2 \\
$D_i$ & diffusion coefficients, i=1, 2, 12. \\
$f_b$ & bulk free energy density function \\
$f_s$ & surface free energy density function \\
$g_i$ & molecule-molecule interaction strength of each components near by the wall \\
$h_i$ & interaction strength proportional to the component volume fractions, i=1, 2. \\
% $J_i$ & mass flux, i=1, 2. \\
$M_i$ & mobility, i=1, 2, 12. \\
% $q_1$ & total free energy in a quadratic form \\
T & temperature of the isotropic system \\
\hline
\end{tabular}
\end{table}

\bibliographystyle{plain}
\bibliography{ms}

% \appendix

\end{document}